# Quantitative Prediction on the Enantioselectivity of Multiple Chiral Iodoarene Scaffolds Based on Whole Geometry

Prema Dhorma Lama[a,†], Surendra Kumar[a,†], Kang Kim[a], Sangjin Ahn[b], Mi-hyun Kim[a,*]

[a]Gachon Institute of Pharmaceutical Science & Department of Pharmacy, College of Pharmacy, Gachon University, 191 Hambakmoeiro, Yeonsu-gu, Incheon, Republic of Korea, [b]Department of Financial Engineering, College of Business, Ajou University, Suwon, 16499, Republic of Korea

This work is dedicated to the researchers who gave their useful comments.

[†] The authors are co-first authors
[*]Author for correspondence
E-mail: kmh0515@gachon.ac.kr

**ABSTRACT**

The mechanistic underpinnings of asymmetric catalysis at atomic levels provide shortcuts for developing the potential value of chiral catalysts beyond the current state-of-the-art. In the enantioselective redox transformations of chiral hypervalent iodines, the present intuition-driven studies require a systematic approach to support their intuitive idea. Arguably, the most practical systematic approach would be based on the reliable quantitative structure-selectivity relationship (QSSR) of diverse and dissimilar chiral scaffolds in an optimal feature space that is universally applied to reactions. Here, we introduce a predictive workflow for the extension of the reaction scope of chiral catalysts across name reactions. For this purpose, whole geometry descriptors were encoded from DFT optimized 3D structures of multiple catalyst scaffolds (113 catalysts in 9 clusters). The molecular descriptors were verified by the statistical comparison of the enantioselective predictive models (classifiers) built from each descriptors of chiral iodoarenes. More notably, capturing the whole molecular geometry through one hot encoding of split three-dimensional molecular fingerprints presented reliable enantioselective predictive models (regressors) for three different name reactions (Kita oxidative spirolactonization, cross coupling, and para-hydroxylative dearomatization) by recycling the data and metadata obtained across reactions. The "potential use value" of this workflow and the advantages of recyclability, compatibility, and generality proved that the workflow can be applied for name reactions other than the aforementioned name reactions (out of samples). Furthermore, for the consensus prediction of ensemble models, this global descriptor can be compared with sterimol parameters (steric descriptor) and noncovalent interaction vectors (stereoelectronic descriptor). This study is one case showing how to overcome the sparsity of experimental data in organic reactions, especially asymmetric catalysis.

***Keywords:*** enantioselective oxidative transformation, asymmetric catalysis, metal-free, chiral catalyst, symmetry, machine learning, quantum mechanics, molecular descriptors, molecular representation

**Main**

In the latest century history of synthetic organic chemistry, asymmetric synthesis has been an important evolution of chemical transformations [1]. Asymmetric catalysis is conceptually outstanding in terms of sustainability, especially, atom economy, among known asymmetric synthetic methods [2-3]. Furthermore, despite their less efficiency than metal catalysts, benign chiral organocatalysts have been intensively developed with advanced ideas and also could be integrated by complex reactions for their applications [2, 4]. Accordingly, mechanistic underpinnings of asymmetric organocatalysis in atomic level can give clues to improve current efficiency of organocatalysis such as reaction scope and selectivity [5-7]. In particular, machine learning (ML) based quantitative elucidation on molecular recognition [8-14] can practically contribute to the design of a new catalyst with powerful high throughput screening.

Arguably, chiral hypervalent iodines (more desirably, hypercoordinate iodines) in Figure 1A have contributed to notable enantio-controlled oxidative reactions with their future-oriented strategy for a sustainable world [15-18]. In other words, while they show metal-like behaviors and functions with their 3c-4e bond, chiral iodoarenes can provide mild metal-free green chemistry with low cost and benign byproducts. Despite these green chemical transformations, their development dominantly relies on researchers' qualitative intuitions rather than quantitative methods [18]. A notable impediment of these computations is owing to the large number of cases resulting from their complex structures, in addition to known general difficulties of catalyst computation. The chiral scaffolds of these catalysts are diverse and dissimilar, which show central chirality [19-20], axial chirality [21-22], planar chirality [23], and helical chirality [24]. Ligands of chiral iodoarene catalysts are determined respectively according to reaction conditions such as a solvent, a co-oxidant and an additive of each reaction through ligand exchanges after oxidation of precatalysts [18, 24, 25]. Thus, almost empirical reports on these reactions did not explicitly describe exact ligands of active catalysts. Furthermore, even if unclear ligands can be unified by one group (eg. MeO-) during general modeling of multiple scaffolds, the unified iodoarene (III) catalysts cannot guarantee geometrical relevance (with their real structures) due to the change of geometry resulting from steric hindrance between L-I-L and substituents of aryl group [25-26]. Therefore, to our knowledge, a quantitative model (either fitted descriptive or validated predictive) on chiral iodoarene catalysts was not reported until our study despite impressive DFT based mechanistic studies [27-30]. Exceptionally, after starting our study, Jacobsen and Sigman's groups reported linear free energy relationship (LFER) models of Ishihara-type catalysts [31]. Their models were built from hypothesized π-π-interaction between catalyst arenes (or substrate arenes) and benzene probe, a surrogate [32]. Thus, the LFER resulted in the substituent optimization of a specific reaction using one chosen chiral scaffold.

More notably, any study did not pragmatically and successfully compare molecular features of multiple catalyst scaffolds (chemotypes) under identical descriptors (variables) [33]. Thus, accumulated mechanistic studies of organocatalysis need to be integrated and systematically analyzed through universe features (presented by numerical variables) to pioneer inscrutable patterns by human intuition [8- 10, 13, 14]. During these analysis, feature spaces, consisting of numerical variables, can be searched and spontaneously manipulated by feature selection algorithms [34] or manifold learning methods [35] to achieve the optimal space, which can lead to reliable quantitative structure-selectivity relationship [36-38].

In this study, to numerically describe the enantioselective transforming potency of chiral catalysts across name reactions in their optimal space, we introduce a predictive workflow for reaction scope extension of chiral catalysts across name reactions (Figure 1B). For this purpose, experimental data of 9 clustered 113 catalysts, of diverse and dissimilar multiple scaffolds (Figure 1C and Supplementary S.Table1), were collected with respective information on the names of their tested reactions, substrates, and reaction conditions in literature. Molecular features of the chiral catalysts were quantified by three type descriptors: (1) sterimol parameters (steric descriptor), (2) three dimensional (3D) radius molecular fingerprint (full geometry descriptor) [39] and non-covalent interaction (NCI) feature vectors (electronic descriptor), which were encoded from density functional theory (DFT) calculated geometry. Then, enantioselective classifiers and enantioselective regressors were built for different name reactions in Figure 1D, using feature selection and machine learning methods. Furthermore, this workflow considered reaction space, catalyst space (of multiple scaffolds), and substrate space. These predictive models of C1 or C2 symmetric iodoarene (III) catalyst scaffolds guides inscrutable pattern by human recognition under this workflow (Figure 1B).

[Insert Fig. 1 here]

**Fig.1: Prediction workflow for asymmetric catalysis across name reactions.**

**Figure 1A.**

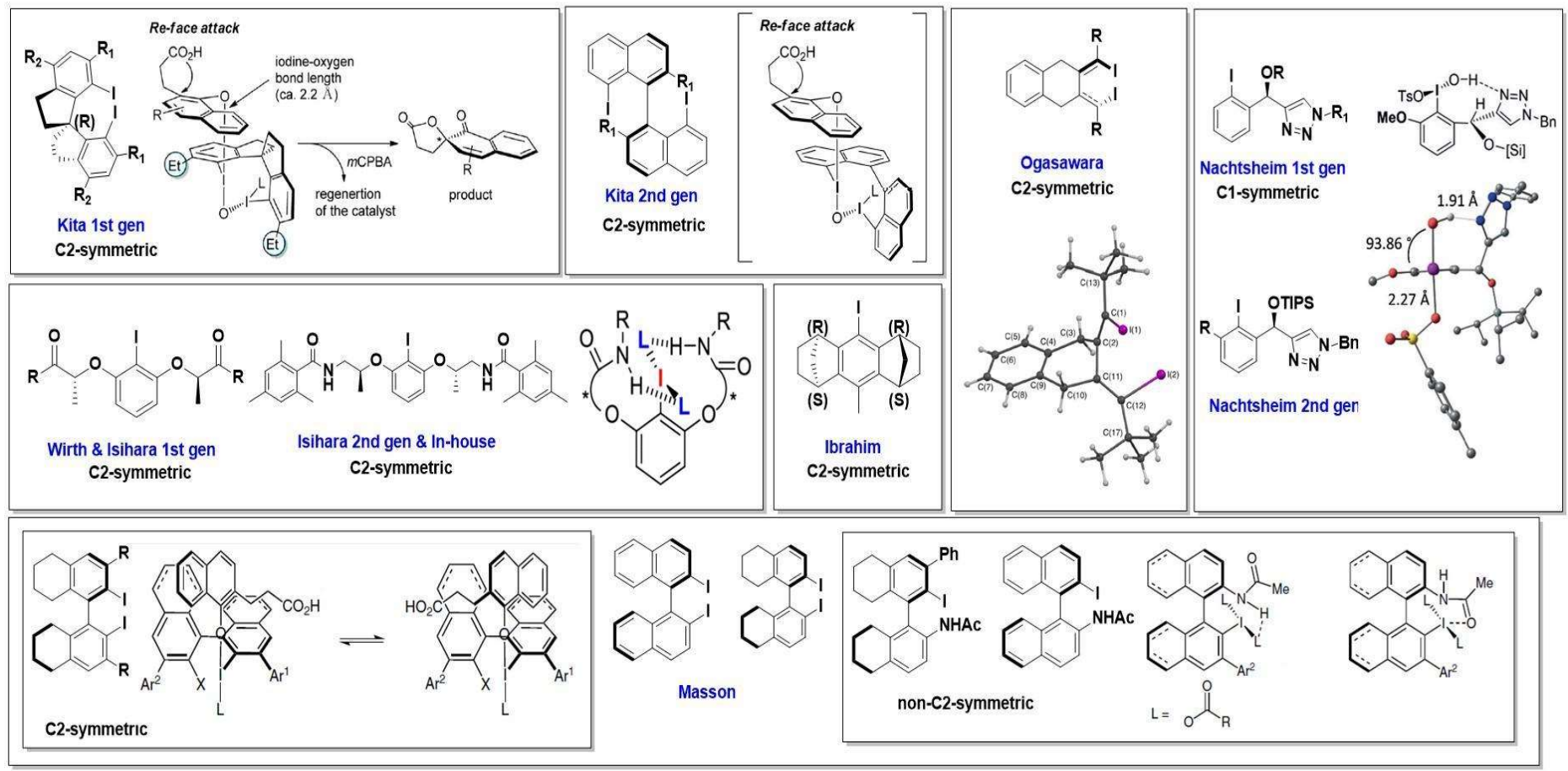

**Figure 1B.**

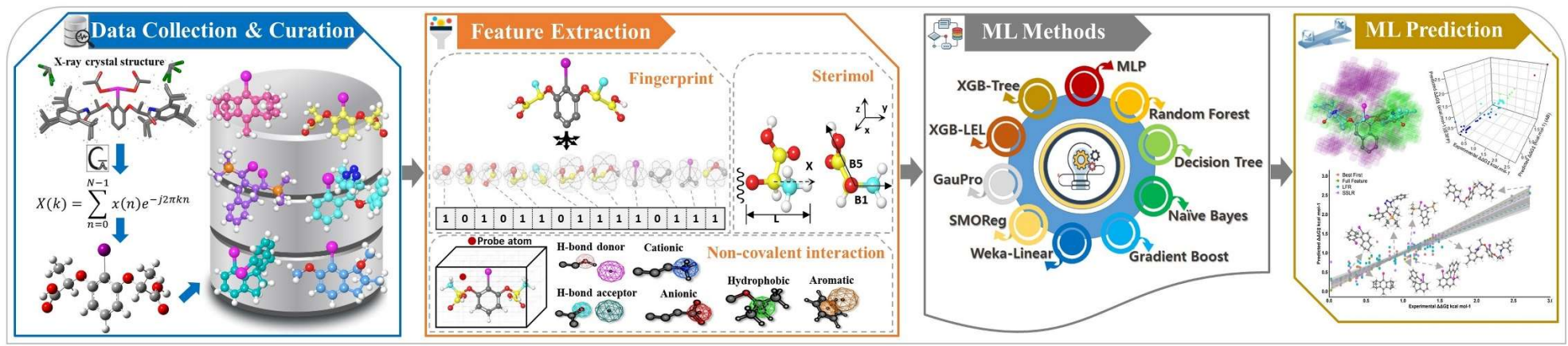

**Figure 1C.**

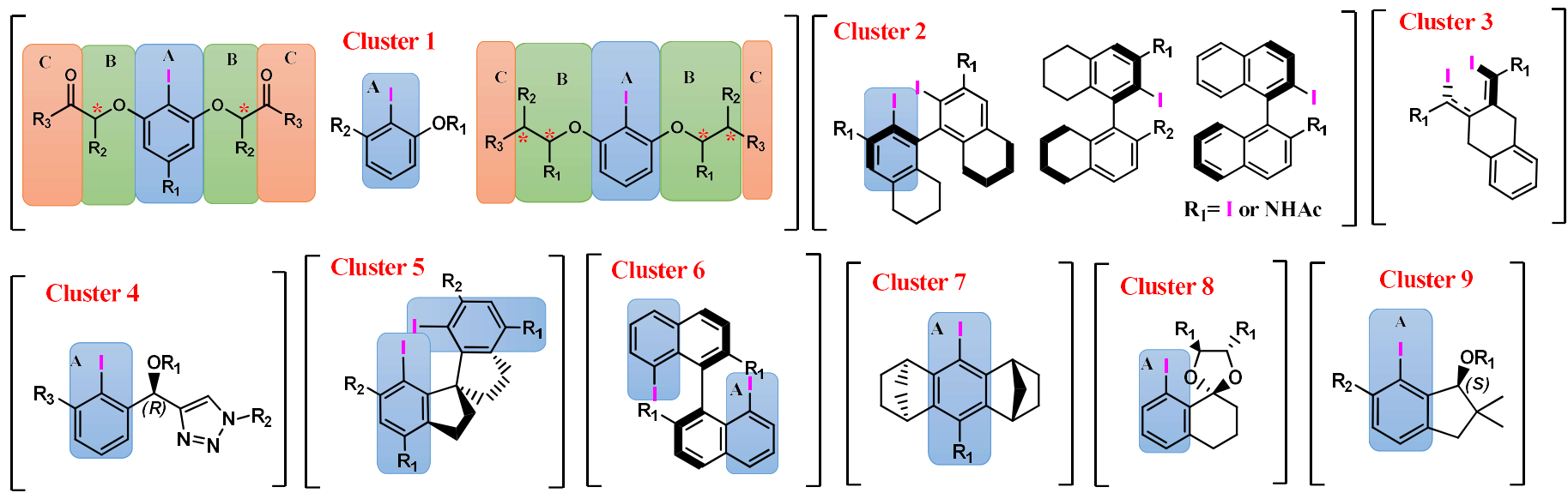

**Figure 1D.**

**Enantioselective oxidative spirolactonization (Dataset 1)**

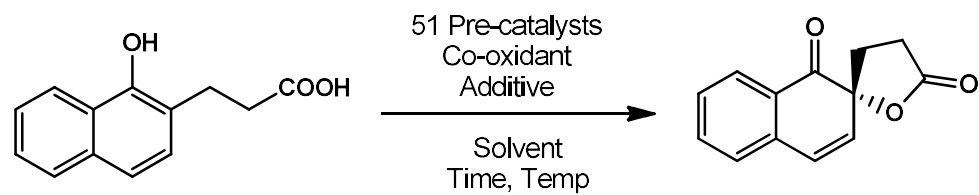

**Enantioselective oxidative cross coupling (Dataset 2)**

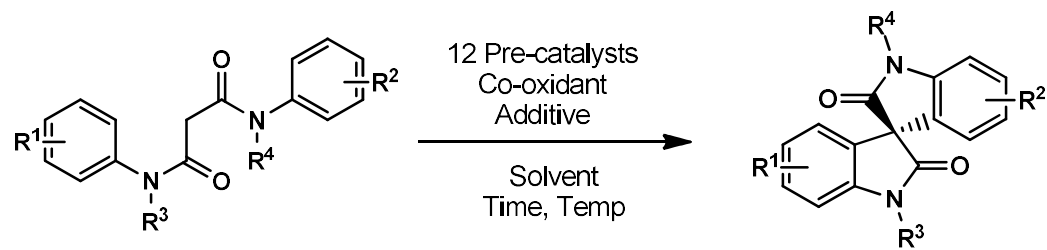

**Enantioselective oxidative *para* dearomatization (Dataset 3)**

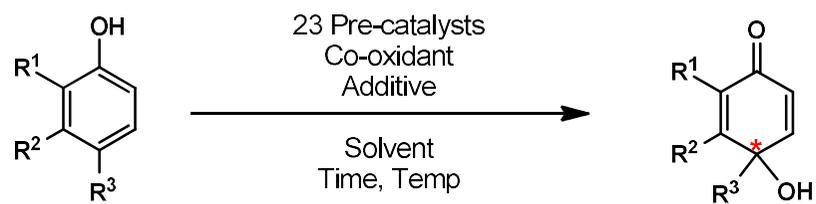

**Figure 1. Prediction workflow for asymmetric catalysis across name reactions.**

 **(A)** Overall dataset (chiral iodoarene pre-catalysts) for workflow development. **(B)** The prediction workflow for enantioselective oxidative transformations of diverse dissimilar C1 or C2 chiral hypervalent iodines scaffolds. Enantioselective predictive models can be built through one hot encoding of DFT optimized geometry. Machine-learning methods realized classification and regression models approximate to the ideal target function correlating encoded descriptors (independent variables) and enantioselectivity (dependent variable). The encoded descriptors with their recyclability, compatibility, and generality were applicable for different name reactions. **(C)** Clustering result of pre-catalysts based on maximum common substructures. **(D)** Name reactions of each dataset: Kita oxidative spirolactonization, cross coupling, and *para* hydroxylative dearomatization.

## Results and Discussion

**Comparison between state-of-the art DFT based ML methods and this work.** With our respect, we could find current state-of-the arts in DFT based prediction of enantioselective organocatalysis: (1) Denmark's grid-based ML model of large scale dataset [40], (2) Sunoj's ML based predictive model of enantioselective catalytic hydrogenation [41], and (3) Sigman's liner regression models of various catalytic reactions [33, 42-44]. In terms of a dataset, while Denmark's and Sunoj's studies used the combination set of a large number of catalysts (1,1'-binaphthyl scaffold) and diverse substrates/reagents, Sigman's studies tended to suggest the best models from a fixed substrate/catalyst with competent descriptors based on deep insight. Recent Sigman's large scale study also follows such large data trend, how to collect data with respective catalysts, reagents, and substrates [44]. Notably, this work focused on the diversity of chiral scaffolds for organocatalysis. Thus, we selected smaller but more diverse dataset rather than those studies. This work collected 9 different catalyst scaffolds (of 113 catalysts) for three different reactions (51, 12, and 22 datasets extracted from the sparse data) from literature and in-house. The state-of-the arts used only one catalyst scaffold but their data showed three dimensions, which consists of > 1000 m*n*r with some vacancy from catalysts or ligands (m), substrates (n), and reagent (r) such as 43 catalysts * 16 substrates*4 reagents of Denmark, 58 ligands * 190 substrates for Sunoj [40-41].

Although all studies internally and externally validated their models, the binominal classification model (of Sunoj et al.) built from large dataset is expected to show more broad applicability domain [41] than non-linear or linear regression models (Sigman et al.). On the other hands, Sigman's interpretable regressors showed high predictive power on local nearest structures based on best fitting of free energy values [33, 42]. However, the state-of-the arts didn't properly present ranking power of their prediction and feature selection algorithm of their models. Thus, this work followed reported validation method and statistical values but added these points. Moreover, general utilities of descriptors were compared in both classification and regression models in this work.

In terms of descriptors, while Sunoj and Sigman used fully QM (quantum mechanical) descriptors from one optimal geometry, Denmark mainly used ASO (average steric occupancy), a molecular interaction field based descriptor to achieve efficient regressors from multiple conformers [40]. While QM descriptors could not directly capture or compare whole geometry of a chiral scaffold, ASO could capture whole geometry and generate numeric values. However, such grid based descriptor should be defined in pairs between each atom of an input molecule and atomic probe of each grid. Thus, ASO is very sensitive to alignment between input molecules. In other words, when chiral catalysts share a limited common substructure, relative position of unshared substructures can make large deviation of ASO descriptors for respective one conformer. The drawback capturing whole geometry made us investigate new descriptors in this study.

**DFT calculation for optimal geometry.** The large of degree of freedom in input structures and number of cases of reaction conditions essentially push us to find a gold triangle between computing cost, speed, and quality. For this purpose, we selected several criteria based on literature. Firstly, we generated molecular descriptors from only catalyst structures and fixed a substrate (eg. phenolic) and a reaction (eg. oxidative dearomatization [45-51]) within one model to elucidate global chiral environments of C1/C2 symmetric diverse scaffolds. [18, 24,

53-55] Thus, one model means one specific reaction such as oxidative dearomatization of 1-naphthol (Figure 1D) [26, 45, 48]. Secondly, empirical data (CCDC x-ray structure) were maximally used as input data for DFT calculation [24, 53-55]. Thirdly, the optimized geometry of catalysts (or precatalysts) were only calculated rather than transition states of reaction complex. In addition, double zeta level of theory was selected for our DFT calculation to follow literature [40-44, 56]. Finally, every molecular descriptor was generated from quantum chemically derived geometry.

Under the chosen criteria, we obtained the optimal geometry of every iodoarenes (Figure 1A and 1C) at double zeta level of theory, M06-2X/6-31G++(d,p) except for iodine atom (I: LanL2DZ) such as Denmark's group (B3LYP/6-31G*)[40], Sigman's group (M06-2X/6–31G** or ωB97XD/6-31G(d)) [42-43], or Sunoj's group (M06-2X/6–31G**) [41]. Each geometry was validated through (1) root mean square distance (RMSD) comparison with available X-ray structures, and (2) vibrational frequency calculations to check the absence of imaginary frequencies. And then, three type molecular descriptors were generated from the optimized geometry: (1) sterimol steric parameters [42], a privileged QM descriptor, for comparison, (2) bit vectors of radial geometry, non molecular pairwise and non molecular interaction field (MIF) [40], and (3) NCI feature vectors.

**Correlation analysis between catalysts and pre-catalysts as catalyst surrogates.** Herein, despite contribution of the ligands to enantioselectivity, we selected precatalysts as surrogates of multiple scaffold catalysts and reduced computing burden to get high speed. To justify this idea, Spearman rank correlation analysis [57] of precatalyst and catalyst (MeO ligands) pairs was performed based on sterimol steric parameters calculated from their optimized geometry in Figure 2B, Supplementary S.Figure1 and Supplementary S.Table2. In detail, representative fifteen catalysts and their precatalysts were chosen from nine clusters (Supplementary S.Table 2). Sterimol using Corey–Pauling–Koltun (CPK) molecular representation captured the dimensions of each chosen substituent along with its bond axis: (1) two width parameters (B1 and B5: the minimum and maximum width orthogonal to a bond axis) and (2) length parameter (L: the length of the substituent along a bond axis) with the defined bond axes (red colored eight bonds) as shown in Figure 2A. Obviously, high correlation between precatalyst and catalysts could justify the suitability of these surrogates. Notably, because some catalysts have C2-symmetry and others have non C2-symmetry catalysts, how to get the best alignment was an important problem. Thus, the shuffling between left-right side also was conducted to know whether or not to make an effect on correlation coefficients (alignment 2 of Figure 2B). In addition, the effect of outlier on correlation coefficient also was checked through their exclusion (alignment 3 of Figure 2B). Furthermore, the frequent distribution of respective sterimol parameters were compared (Figure2C). Three distributions of precatalysts, catalysts, and their combined were similar and their correlation coefficients between sterimol pairs were similar.

[Insert Fig. 2 here]

**Fig2.: Correlation analysis of steric features between catalysts and precatalysts.**

**Figure 2A.**

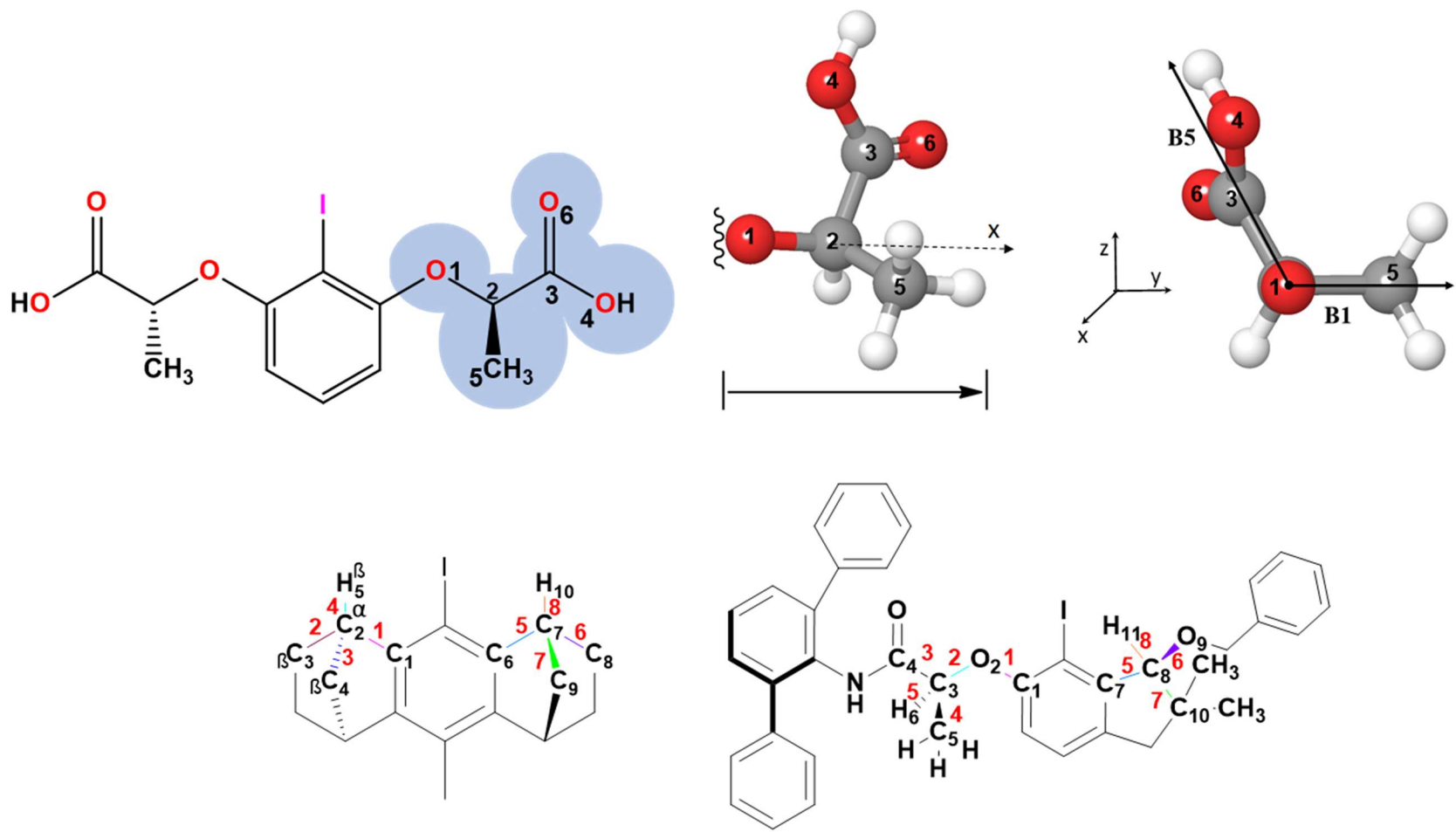

**Figure 2B.**

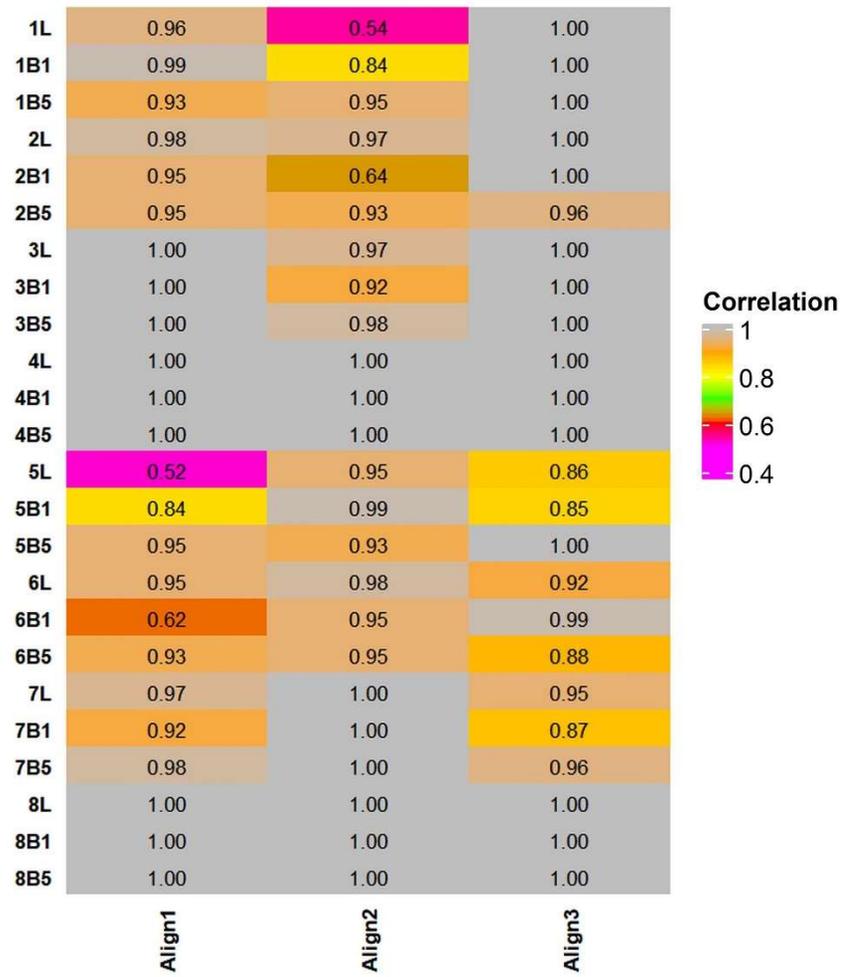

**Figure 2C.**

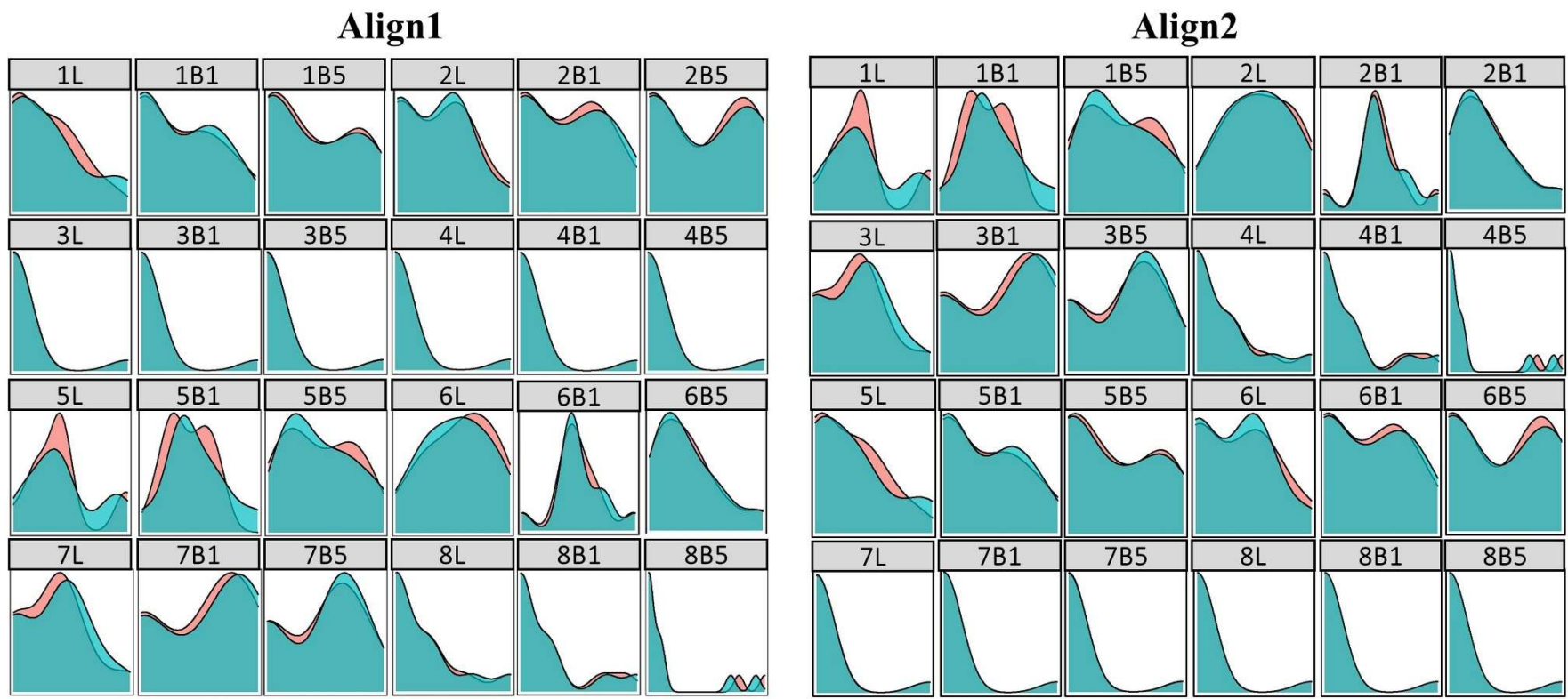

**Figure 2. Correlation analysis of steric features between catalysts and precatalysts.**

**(A)** Definition of sterimol parameters chosen in this study. In order to capture steric features from thirteen catalysts and their precatalysts, eight bond axes (red colored) were chosen.

The first digit of every sterimol parameters is identical to defined bon axis number. **(B)** Under three alignments, identical sterimol parameter values were compared with correlation coefficients between the fifteen catalysts and their precatalysts. **(C)** The frequent distribution of respective sterimol values were visualized from the precatalysts and their catalysts. Red colored distributions are precatalysts (abbrev. pcat) and blue green colored dots and distributions are catalysts (abbrev. tcat).

**An alignment tolerable 3D molecular fingerprint as a whole geometric descriptor.** In principle, a steric descriptor on whole geometry was required to elucidate global chiral environments of diverse C1/C2 scaffolds showing structural and numerical variance. Sterimol parameters are frequently used steric descriptor for DFT based models to prove their competency in reported descriptive and predictive models [38, 56]. However, the sterimol parameters cannot simultaneously describe whole geometry and they cannot be automatically generated due to requiring axis definition of respective compound (Figure 2A). In the case of MIF, grid based descriptor, MIFs can simultaneously generate whole geometry descriptors through the calculation between atoms of input structures and atomic probes (eg. sp3 carbon cation of CoMFA, sp3 carbon of ASO) in grids. Thus, their accuracy critically depends on the alignment of compounds (dataset) as well as 3D conformations [40]. More seriously, when core structures of dataset are dissimilar such as our nine chiral scaffolds, perfect superimposition is impossible to capture structure information (atom-probe pair) accurately from infinite occasions in grid space. Furthermore, grid-independent MIF-based methods were reported with their eligibility for diverse scaffolds but it also was reported that these methods show underperformance with respect to their alignment dependent counterparts [38].

To solve this drawback, we searched 3D molecular descriptors and excluded pairwise methods [39, 58, 59]. And then, Keiser's 3D molecular fingerprint, extended three-dimensional fingerprint (E3FP) [39], was selected because it could globally capture geometrical relation of all atoms, from input structures, without alignment (superimposition). 3D Cartesian coordinate of every atom in each iodoarenes could be transformed into binary identifiers through iterative shell radius multipliers (*L: level* and *r: radius* parameters) [39]. Regardless of covalent bonds or not between atoms, these iterative shells capture the 3D atom neighborhood patterns within an input structure (in this study, DFT calculated geometry). Because the iterative multipliers make cascade change of identifiers (32-bit -> folded -> bitvector), human recognizable information cannot remain in the final identifier and E3FPs of molecules cannot be decoded to their initial structures. However, obviously, different conformers of one molecule produced different unique E3FPs to prove uniqueness of the identifier (Supplementary. S.Figure2). In addition, the E3FP developer, Keiser, also reported multiple E3FPs of one molecule respective to multiple conformers. Furthermore, the 3D fingerprint permitted simultaneous and automatic geometrical description on both local and global environments of chiral iodoarenes. More notably, the 3D fingerprint showed alignment tolerable representation of input conformers due to independency of the encoding between molecules. In order, E3FPs as objects in Python could be converted to writable numeric data (one hot encoding of 1024 variables).

To ensure that the E3FP bitvectors/bitstrings (modified from objects to writable data of 1024 variables, which was one-hot encoded after bit splitting) subtly describe structural features pertaining to enantio-control, E3FP based classification models were built under versatile conditions including (1) Random Forest (RF) [60], (2) Decision Tree (DT) [61], (3) Naïve Bayes (NB) [62], (4) Gradient Boost Tree (GB) [63], (5) Xtra-Gradient Boost Linear (XGB-Linear), (6) Xtra-Gradient Boost Tree (XGB-Tree) [64], (7) Support Vector Machine (SMOReg) [65], and (8) Multi-Layer Perceptron (MLP) (Figure 3). The robustness and competency of the fingerprint based classifiers were tested according to (1) feature optimization [66], and (2) classification algorithms (Supplementary S.Table3), (3) threshold between active (more selective) and inactive (less selective), (4) dividing methods of training set (TR) and test set (TS) with 10 times unbiased sampling, and (5) validation methods

(internal n-fold cross validation & external validation using TS) showing statistical performance of all trials. Feature selection methods on whole dataset gave 39 variables (3.8%) and 176 variables (17.2%) respectively from Best First (BF) and High-Correlation Filter (HCF) approach. Notably, these fingerprints were tolerable for feature reduction (loss of some variables after selection) and were highly economic than ASO of Denmark group [40]. While 39 variables were used for our enantioselective prediction (up to 3.8 % of E3FP bit vectors), 16,384 variables were used for ASO based prediction in literature. Among a total of 240 ML models (3 selectors * 8 classifiers *10 seed states), the classification metrics for BF based ML models yields higher performance in terms of train, test set and valid set with 5-Fold cross-validation metrics. Distinctly, BF-NB, Best-First as feature selector and NB as classifier (MCC = 0.943±0.040, Accuracy = 0.973±0.019) was superior to Full-feature-MLP (MCC = 0.475±0.064, Accuracy = 0.753±0.030), the best of full feature models, and HCF-SVM (MCC = 0.440±0.093, Accuracy = 0.733±0.044), the best of HCF feature selection models in Figure 3A and Supplementary S.Table 3. Moreover, these more reliable BF based models meet the principles of parsimony (simple model) as well as their robustness.

Similarly, for sterimol descriptors, the selected best average model is shown in Figure 3B and Supplementary S.Table 4. When models with MCC > 0.4 can be considered as a predictive model, after internal (of 5-fold CV) and external validation (of test set), these were barely predictive: Full Features-MLP (MCC = 0.607±0.194 (test), 0.519±0.087 (5-FoldCV); Accuracy = 0.791±0.096 (test), 0.773±0.042 (5-FoldCV)), HCF-MLP (MCC = 0.507±0.400 (test), 0.501±0.059 (5-FoldCV), Accuracy = 0.764±0.197 (test), 0.767±0.027 (5-FoldCV)) and BF-SVM (MCC = 0.539±0.116 (test), 0.448±0.087 (5-FoldCV), Accuracy = 0.736±0.080 (test), 0.722±0.041 (5-FoldCV)) in Figure 3B and 3D. After comparing all the classification metrics, full Feature-based MLP emerges was the best classifier. Despite similar number of variables, Best First (4 variables) and HCF (5 variables) showed a huge difference in predictability on the train set. Upon analyzing the type of features selected by the BF and HCF feature, only two common features (3L, 8L) were common to show different thresholds of filtering criteria in these feature selection methods. Moreover, high correlation coefficients between different sterimol parameters (Figure 2C) and decreased performance after feature selection makes us suspect collinearity issue between sterimol parameters.

More notably, we could clearly conclude the maximal statistical performance of sterimol based classifiers is inferior to E3FP based classifiers (Figure3 (A) vs (B), (C) vs (D)). Furthermore, only E3FP based classifiers permitted comparative or enhanced performance after feature reduction (Figure 3). It suggests that 3D fingerprint, a 3D molecular representation, can provide smart description for diverse and dissimilar catalysts and permit feature optimization by machine. Most notably, our results suggest while sterimol parameters are confined to substructure of these catalysts, E3FP can capture the concrete information on chiral environment through describing whole geometry. In other words, despite their heterogeneity between iodoarene scaffolds, it seems that the global descriptor (E3FP) didn't fail to detect delicate steric information of the scaffold and at least better detect the information rather than the local descriptor (Sterimol).

[Insert Fig. 3 here]

**Fig3.: Competency comparison of molecular descriptors.**

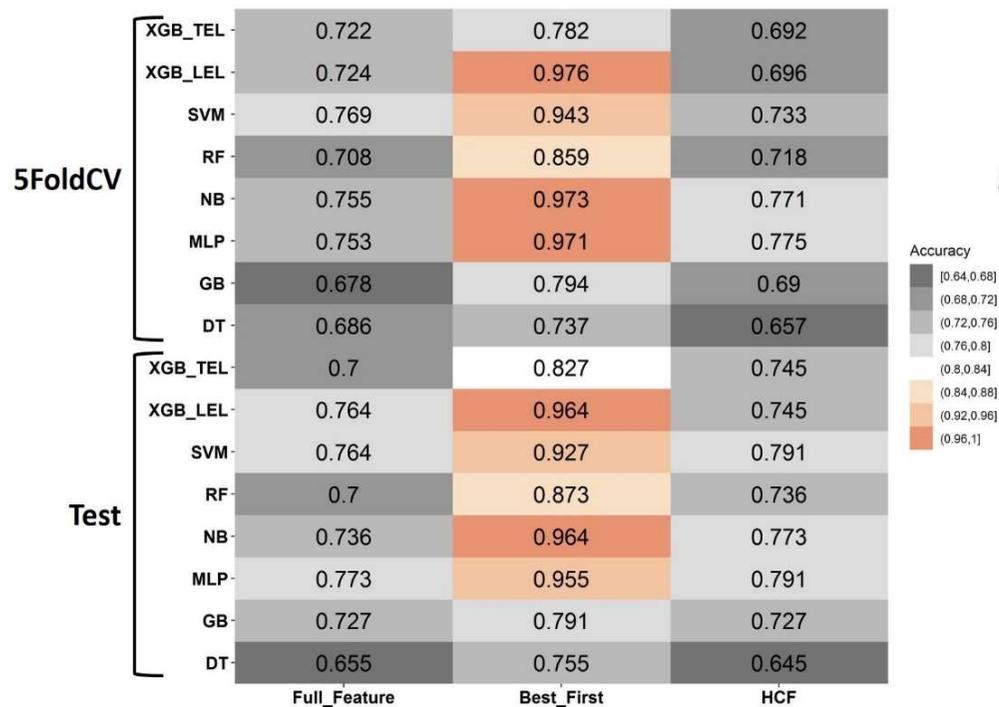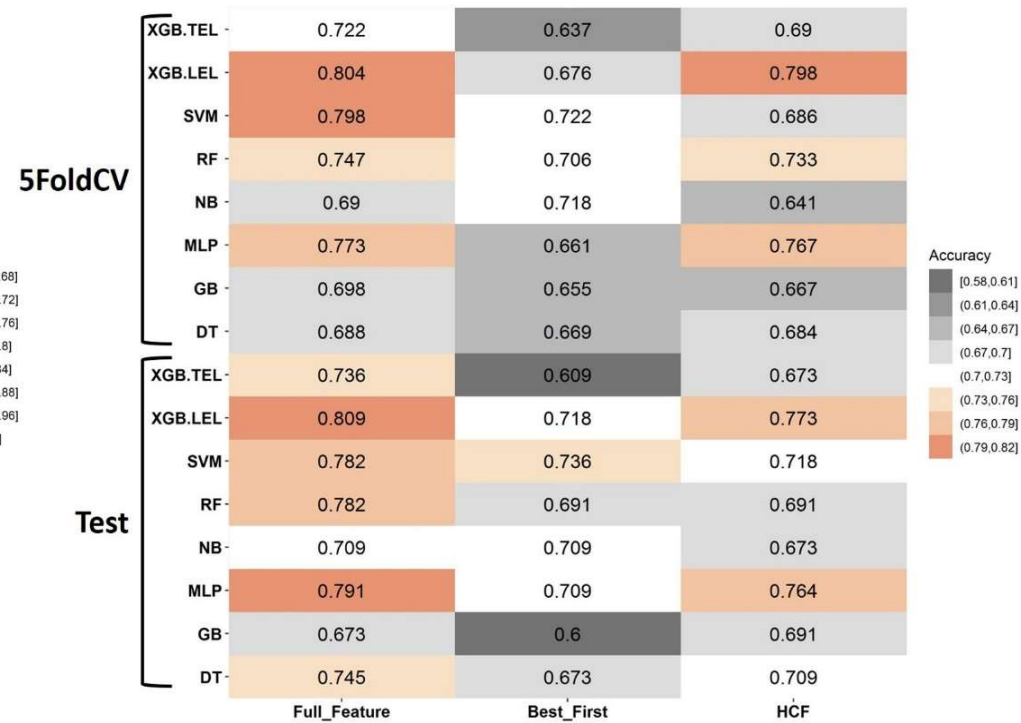

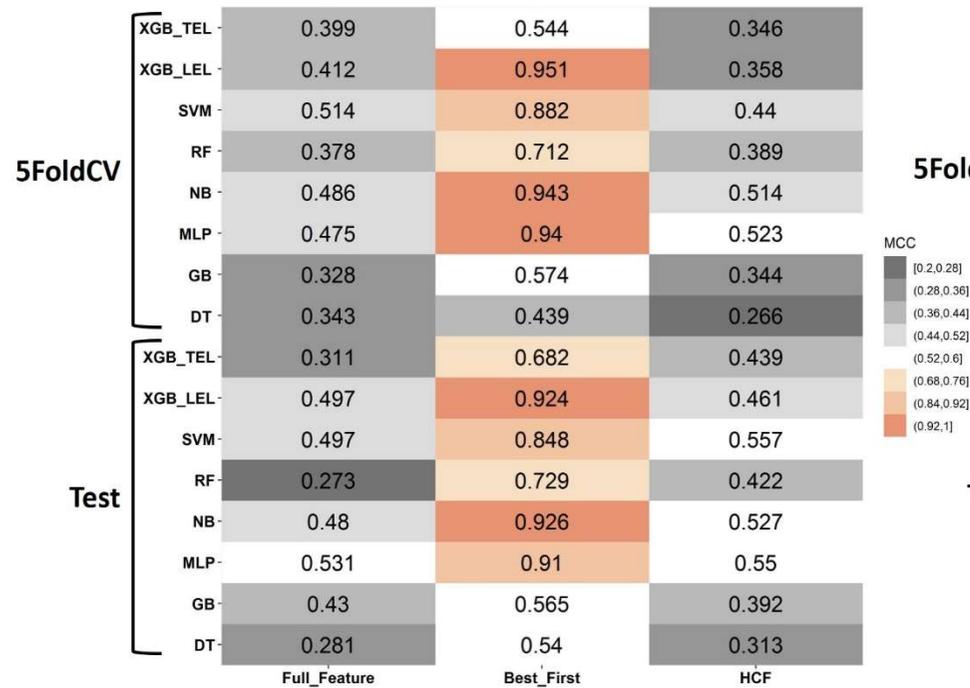
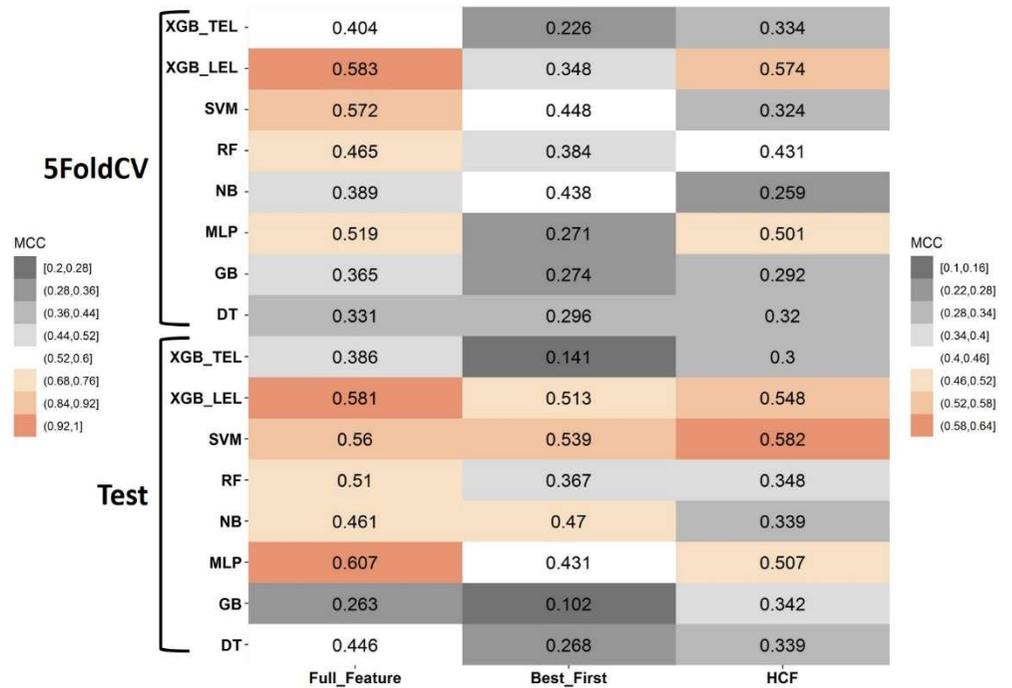

**Figure 3. Competency comparison of molecular descriptors.**

Performance comparison between 3D fingerprint (E3FP) or sterimol based classification models. XGB-TEL: Xtra-Gradient Boost Tree Linear, XGB-LEL: Xtra-Gradient Boost Linear, SVM: Support Vector Machine, RF: Random Forest, NB: Naïve Bayes, MLP: Multi-Layer Perceptron, GB: Gradient Boost Tree, and DT: Decision Tree. **(A)** Accuracy of E3FP based models, **(B)** Accuracy of sterimol based models, **(C)** MCC of E3FP based models, and **(D)** MCC of sterimol based models.

**Predictability of the whole geometrical based enantioselective models.** Based on these promising results, scoring power and ranking power [57] were further investigated to evaluate practical usefulness of our current workflow. Promptly, regression models were built with several feature selection algorithms and regression algorithms. To compare performance of various combination of feature selectors-regressors, statistical values were calculated from experimental and predicted ΔΔG‡, which was transformed from enantiomeric excess [31-33, 42-44]. For scoring power, determination coefficients (R-squared) is the most general residual analytic criterion between experiments and predictions. Thus, mean absolute error (MAE), mean squared error (MSE), root-mean-square error (RMSE) and mean absolute deviation (MAD) from training, test, and validation set were calculated together with their determination coefficients. Based on classification result, here we employed RF, XGB-Linear, XGB-Tree, linear, SMOReg[67], GP (Gaussian Process) [68], and MLP algorithms to achieve a more reliable predictive model under three feature selection methods (BF, SSLR, and LFR) [66]. For regression, ranking power, spearman ranking correlation coefficient (Sp) was calculated [57]. To check the robustness of each model, we performed the 5-fold cross-validation with the 10 different random seed numbers. The best model was selected after comparing the RMSE_test, Sp_test, R2_cv, and RMSE_cv. Furthermore, some other parameters, such as RMSE_train, Sp_train, MAD, MAE, MSE were considered, while further model selection.

[Insert Fig. 4 here -2D scatter plot of best regression models]

**Fig 4.: Best enantioselective regression model of Kita oxidative Spirolactonization.**

Obviously, the 3D fingerprint (global descriptor) showed more competent predictive performance than sterimol (local descriptor) in both scoring power and ranking power in Figure 4 and Supplementary. Table 5. Even if SMOReg using full features (1024 variables) had better statistical values on the test set as well as a train set, no model achieved acceptable $R2\_cv$ (> 0.5), which suspect that this model may suffer over-fitting. In the contrast to full feature models, feature selector presented acceptable $R2\_cv$ (> 0.5). SMOReg and GP methods, which use polynomial kernel and former also uses sequential minimal optimization (SMO) algorithm [67], proved better scoring power and ranking power with SSLR (21 variables) and LFR selectors (31 variables) rather than with full features (1024 variables) and Best-First (45 variables) feature selection method. Furthermore, the scoring power was stably and robustly retained during different 10 times running with different seed numbers (Supplementary. Table 5).

Obviously, SSLR-SMOReg model presented the best prediction quality with $R2\_test$ = 0.766, $Sp$ = 0.835, $RMSE$ = 0.205 among every models (Figure 4A and Supplementary. Table 5). In addition, when the scatter plot of SMOReg regressors between prediction and experiments clearly showed consistent residual pattern in four feature selection methods (Figure 4B). LFR-GP models could be the 2[nd] best with the respective statistical values (LFR-GP: $R2\_test$ = 0.668, $Sp$ = 0.753, $RMSE$ = 0.258). In contrast, either SMOReg or GP were not compatibles for sterimol parameters (local descriptor). More notably, either any selector or regressor didn't present reliable predictive result showing $R2\_cv$(> 0.5) to demonstrate the balanced performance in the train, test, and cross-validation to escape any over-fitting issue . Every E3FP-based model showed the higher performance than any sterimol-based model.

Furthermore, while some feature reduction of E3FP improved predictive power of their models, sterimol didn't show improvement for any feature reduction. It suggested the probability that the E3FP based models can work well in general dataset without overfitting under optimization of learners (feature selectors and regressors). Most notably, despite heterogeneous core structures (9 clustered scaffold), MAD existed in the range of 0.187 to 0.423 in every model and the MAD values was a comparative result to the state-of-the art models of Denmark (MIF based universe model for binaphtyl phosphoric acids) [40]. Thus, current result indicate that sterimol descriptor can be important for only congeneric series of catalyst such as Denmarkm Sigman, and Sunojo [39-41], while considering diversity of scaffold such as our present dataset, full geometry descriptors (E3FP) emerges more suitable for enantioselectivity ratio (ΔΔG) of catalysts.

**Figure 4A.**

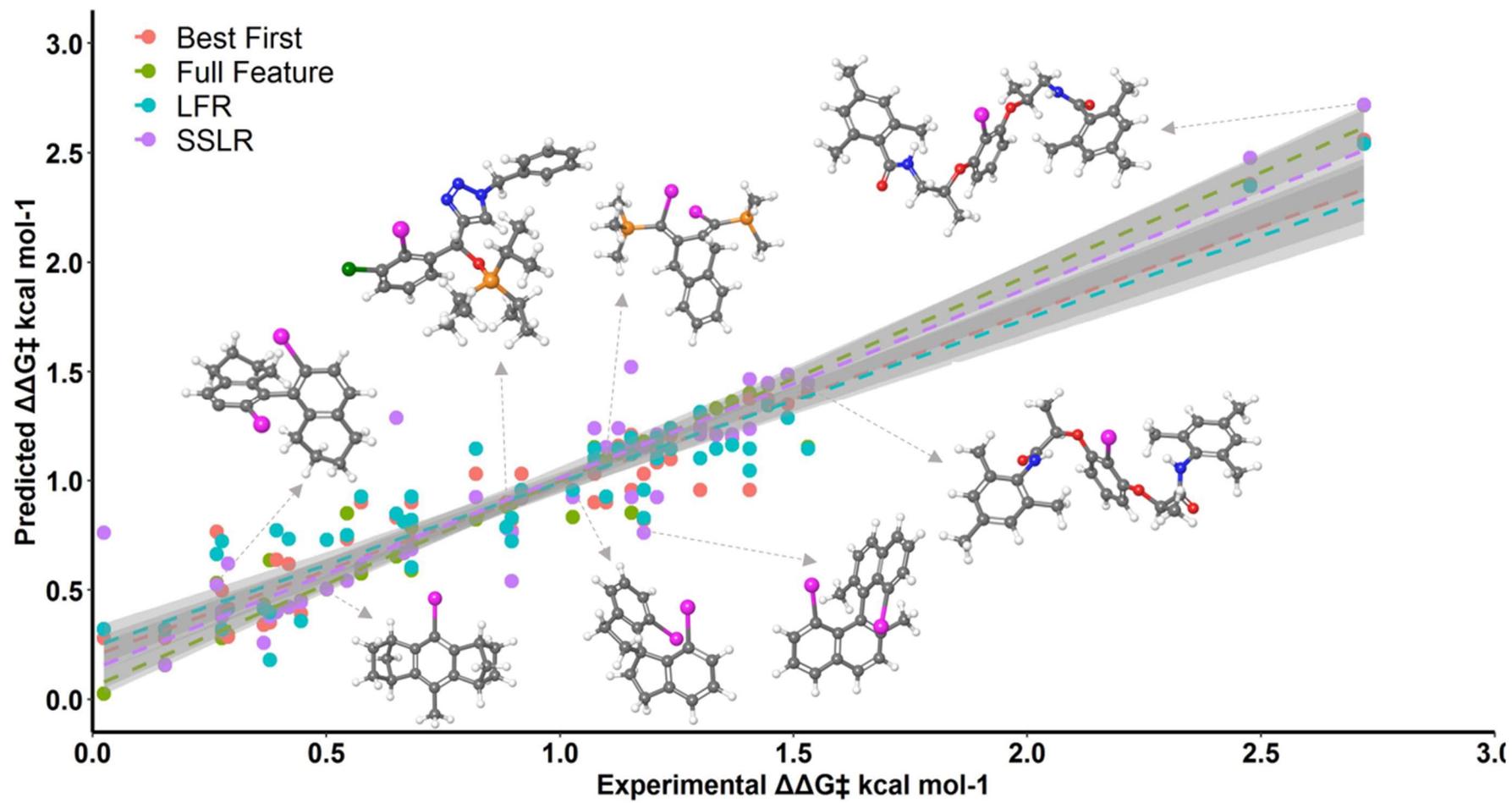

**Figure 4B.**

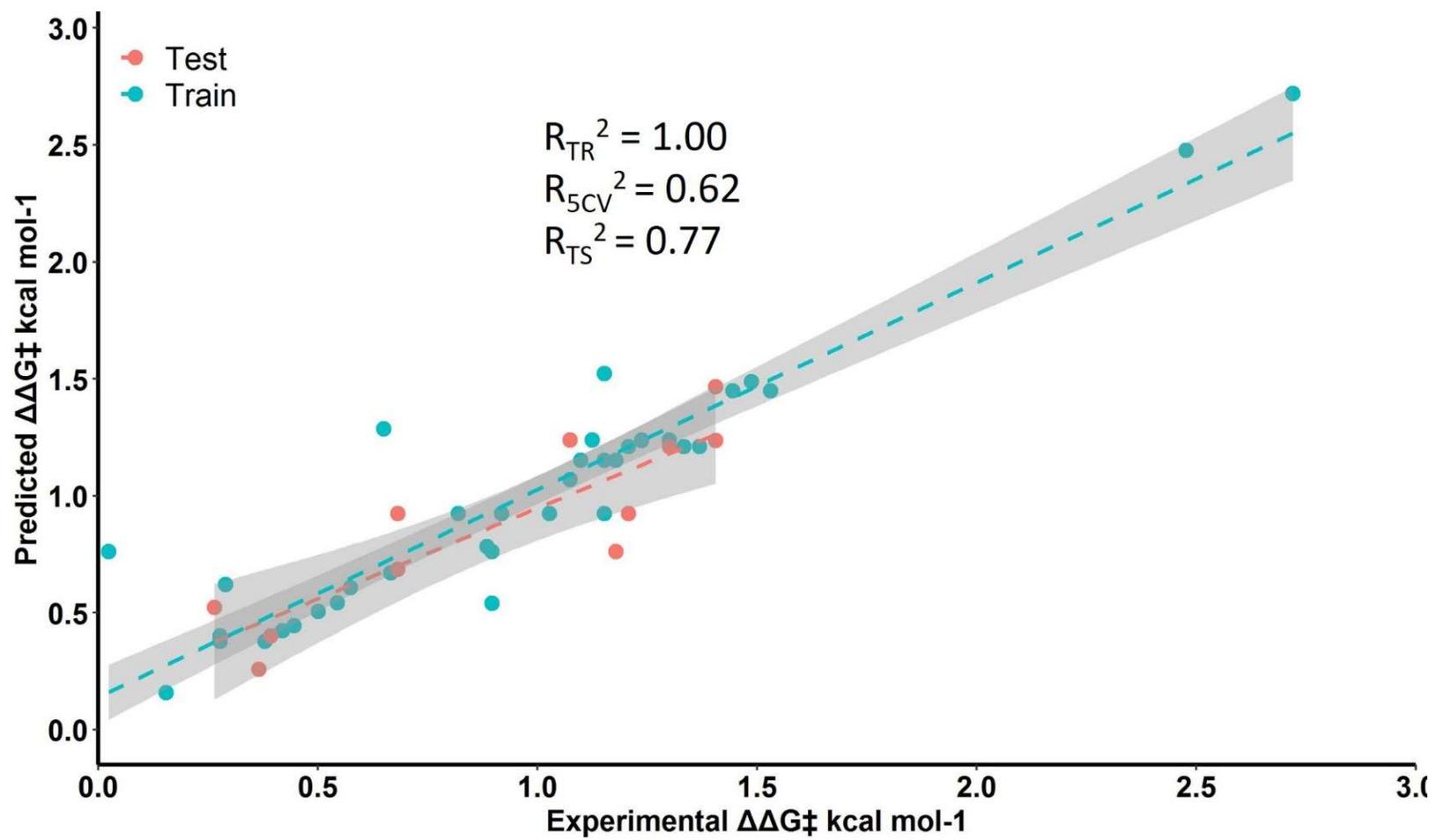

**Figure 4C.**

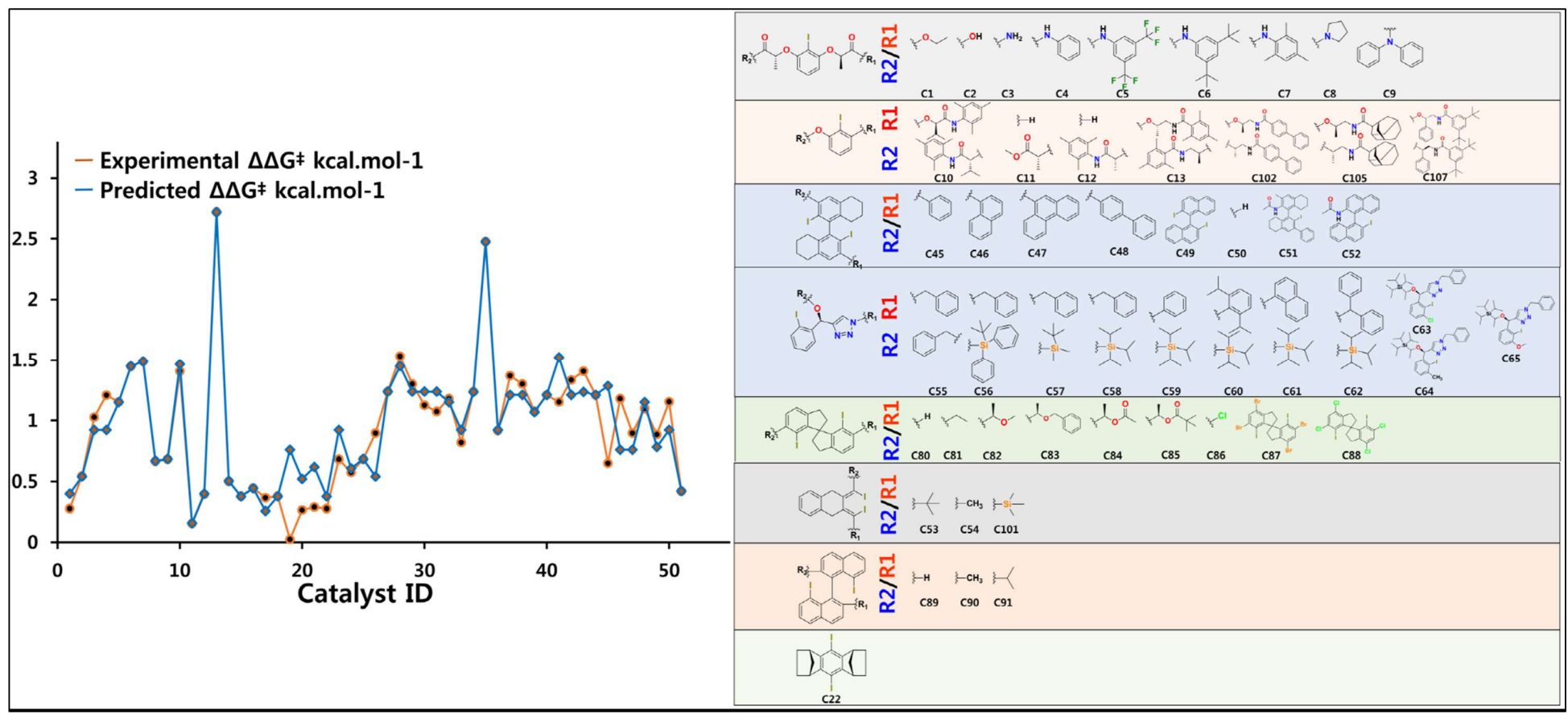

**Figure 4. Best enantioselective regression model of Kita oxidative Spirolactonization.**

Predictive power of 3D fingerprint (E3FP) based regression models. **(A)** The prediction results of regression models according to feature selections (Full, BestFirst, SSLR, and LFR), **(B)** The internal and external validation result of the best model (feature selector: SSLR, regressor: SMOReg), and **(C)** The experimental and predicted $\Delta\Delta G^{\ddagger}$ of dataset1 (No. of Set: 51, x-axis: substituents of catalysts, y-axis: free energy gap calculated from % enantioselectivity).

**Generality of 3D molecular fingerprint based models.** With our delight, the positive results motivated us to further study '*potential use value*' of our current workflow and generality (es., an application into another named reaction or substrate). For this purpose, we prepared the definition of problems: (1) whenever a new asymmetric reaction is reported, whether prior data with these features (descriptors) can be reused with the added new (recycling of metadata), (2) whether or not such metadata also are directly applicable for another name reaction or replaced substrates (compatibility of metadata), and (3) how to analyze these robust models (interpretability of metadata). To solve the defined problems, firstly, the dataset 1 of 51 catalysts (Model1: enantioselective Kita oxidative spirolactonization) was replaced with the dataset 2 of 12 catalysts (Model2A: enantioselective oxidative cross coupling) as another name reaction, and the dataset 3 of 23 catalysts (Model2B: enantioselective *para* hydroxylative dearomatization) as a far different substrate (*ortho* of dataset1 vs *para* of dataset3) of the identical named reaction. In the dataset replacement, some catalysts were reused with the added new from literature and featurization method of added data was identical to dataset 1 (Figure 5A). While dataset 1 contains seven different scaffolds (2 to 98 %ee from cluster 1 to cluster 7), dataset2 consists of only one scaffold (5-83 %ee from cluster 1) and dataset3 consists of two reused scaffolds and two new scaffolds (14-93 %ee from cluster 1, 4, 8, and 9). 2D structures of duplicated catalysts in these three name reactions could be visualized with their selectivity to directly show structure–selectivity trends (Figure 5B). Regardless of the potency differences of catalysts between tested reactions, bitvectors of catalysts C2, C5, and C7 could be recycled to build a different name reaction. As shown in the Figure 5A, every descriptor (meta data) without any manipulation could be directly used to build models of each name reactions. Moreover, the dataset3 could be updated with additional data and used to build models of enantioselective *para* hydroxylative dearomatization. These models were compared before and after updating data.

Secondly, classification and regression models of these additional two datasets were built and internally and externally validated to present robust performance in training, test and validation sets (Figure 5C & D and supplementary S.Figure 3&4). Very surprisingly, the whole geometric descriptor based classification models of dataset 2 and 3 could show their robust performance in the change of a reaction as well as the composition of dataset to show recyclability of E3FP descriptors and their compatibility (supplementary S.Table 6 &7). For example, the Best First based XGB-LEL (Linear Ensemble Learner) of dataset 3 has reliable performance with MCC and AUC on Test Set (0.933; 0.967) and Valid Set (0.983; 0.992). Notably, in all these classifiers, Best First emerges as the best attribute selection method with the minimal number of features representing the overall predictability of ML models. Based on these desirable result, these datasets also were used to further explore the regressors approximate to ideal models (Model 2A and 2B from respective datasets) in Figure 5C & D and supplementary S.Table 8&9. The predictive performance of these models was accessed on a valid set with 5-foldCV statistics. For the dataset 2, when our evaluating them with RMSE, LFR based GP model for oxidative cross coupling showed the best performance (RMSE_test: 0.270; $R^2$_cv: 0.862; $R^2$_RMSE:0.226). Thus, the robustness of such performance could be observed under 10 times with different dataset division (Figure 5C & D and supplementary S.Table8&9). Notably, Best-First based MLP model was selected as the best model (RMSE_test: 0.116; $R^2$_cv: 0.812; R2_RMSE:0.170) after this iterative validation (Figure 5C & D). In terms of data quality and diversity, dataset3 could permit us to find the best regression model more clearly than dataset2. Despite maximal performance for the train set with a full feature set, the RF based model did not perform well on the valid set. The other attribute-based ML model has at par statistics on train set and test set, however, SSLR and LFR based model performed poorly on valid set with higher RMSE. Overall, the Best-First based MLP model for *para* hydroxylative dearomatization has better performance than other models with RMSE_test = 0.116 and $R^2$_cv = 0.812; RMSE_CV:0.170).

Desirably, this workflow based on one hot encoded geometry can suggest us its application across name reactions and compatibility with another catalyst space, another reaction space, and another substrate space. Finally, 3D fingerprints of 1024-bit vector asked us its interpretation. Typical interpretation methods, independent on machine learning algorithms, include feature(variable) importance analysis, sensitivity analysis, partial derivatives approach, and so on [37]. Furthermore, visualization of descriptor contributions such as CoMFA contour map practically helps researcher to catch the perspective meaning of their models. Thus, in this study, we conducted the feature importance analysis based on SHAP (SHapley Additive exPlanations) values of Lundberg and Lee [69]. While E3FPs cannot be directly decoded into 3D conformers or 2D structures of the original input, sterimol parameters of Sigman group can show direct meaning, which molecular feature should be how to be modified. Very fortunately, an E3FP fingerprint is a bit vector, where a specific bit theoretically indicates the presence of a molecular substructure such as SMARTS pattern (string type of 2D molecular representation). Based on this point, the visualized feature importance of the bitvector is useful guidance for synthetic chemicsts.

[Insert Figure5 here]

**Fig 5.: Recyclability, compatibility, and generality of one hot encoded model**

**Non-covalent interaction (NCI) feature vectors as MIF descriptors.** After achieving predictability and generality, two considerations made us continue further investigation. One consideration is the visualization of model, which is human recognition friendly, and another is the consideration on electronic features. Obviously, upper studied models only used steric features. Thus, when we can build reliable models having electronic features, these models can be a part of ensemble prediction to check the consensus with E3FP based prediction. Furthermore, some MIF descriptors of cheminformatics can permit visualization of descriptor contributions with graphic user interface of high quality. Thus, we tried to build MIF model, especially, of electronic features.

For this purpose, we did the benchmarking of pharmacophore modeling as well as 3D quantitative structure-activity relationship (QSAR) in drug discovery. Even though the summation of NCI features in organocatalyst–substrate complexes can be different in terms of scale, weight of specific types, and the number of features with protein-ligand complexes, every complex and their non-covalent interaction follows general chemistry principles explained by the period table of elements. Firstly, six sphere type NCI feature vectors, which consist of hydrogen bonding donor (HD), acceptor (HA), cation (C), anion (A), hydrophobic (H), and ring (R) symbolizing pi-pi or pi-cation interactions, were chosen through pharmacophore modeling tool. Secondly, the sphere type NCI feature vectors could be modified according the sphere size, from one unified size to van der Waals radius of each atom (atom-based NCI feature). Thirdly, field based four type feature vectors, which could be calculated based on grids and probes such as CoMFA and CoMSIA, also were considered. More concretely, Gaussian and modified Gaussian based feature vectors were tested, which the former uses Gaussian function of the distance between a grid point and an atom and the later additionally includes partial charge of each atom. Linear regression models were built from these features in parallel and compared under versatile learning conditions: (1) feature type, (2) the number of features, (3) dataset of different named reactions and substrates, (4) validation methods, and (5) partial linear square (PLS) factor (see the performance of these NCI vector based regression model in supplementary S.Figure5-11). Expectedly, these high dimensional MIF could similarly work such as reported ASO methods with visualized interpretability. Naturally, meta data of these type models could not be recyclable across name reactions due to pairwise dependency of descriptors.

Most notably, the correlation coefficients and determination coefficients (R-squared) suggested us atom-based QSAR model as the best PLS model among these trials. The atom–based model (RMSE_test=0.30 and $R^2$_cv=0.73, $R^2$_cv (scramble) = 0.76) showed the feature contribution order of hydrophobic > electron withdrawing > H-bond donor (hydrophobic: 0.799, electron withdrawing: 0.083, and H-bond donor: 0.032). While almost catalysts have C2 symmetry and showed successful enantioselectivity, C2 symmetry scaffold was less preferred to C1 in the view of hydrophobicity due to asymmetry of the hydrophobic feature (green: preferred hydrophobicity, purple: disfavored hydrophobicity in Figure 6A). It suggests that further structural tuning of C1 symmetry catalyst can be developed beyond Nachtsheim or Maruoka catalysts. Furthermore, $\alpha$, $\beta$, and $\gamma$ position in ortho substituent of iodobenzene preferred electron withdrawing group but this feature has narrow region limited to heteroatom such as *o*-oxygen atoms or nitrogen atom of triazole group or O-silyl group of Nachtsheim catalysts (Figure 6B).

In sequence, the NCI model could be compared with E3FP model and their consensus was checked. Very surprisingly, the atom-based model showed high correlation with both experimental selectivity (Spearman Corr.: 0.90) and E3FP based regression model (Spearman Corr.: 0.86) to prove their consensus in Figure 6C. Furthermore, residual analysis was conducted to compare scoring power of these two models. Despite reliable determination coefficients of every model, the residual analysis presented the limitation of NCI model (Figure 6C and 6D). When comparing with E3FP model, residual deviation is larger than E3FP based regressor (Figure 6D) and the range of predictable free energy was narrower than E3FP-regressor or experimental values (Figure 6C). More notably, the prediction of higher potent catalyst ($\Delta\Delta G^{\ddagger}$ > 2.0 kcal/mol, enantiomeric excess: > 95 ee%) was worse than the prediction of inactive catalysts in NCI model. Thus, NCI model showed limited scoring power and predictability range. Even if the NCI vector based regressor showed these limitation, such useful contour maps for human recognition can suggest the ensemble prediction to compensate interpretability of E3FP based prediction. In other words, we suggest the prediction workflow, which build both E3FP–regressor and NCI-regressor to gain all merits of these models; (1) the ensemble prediction (for consensus), (2) higher scoring power (from E3FP), and (3) visualization of model (from NCI).

[Insert Figure6 here]

**Fig 6.:The consensus prediction of the ensemble model.**

**Figure 5A.**

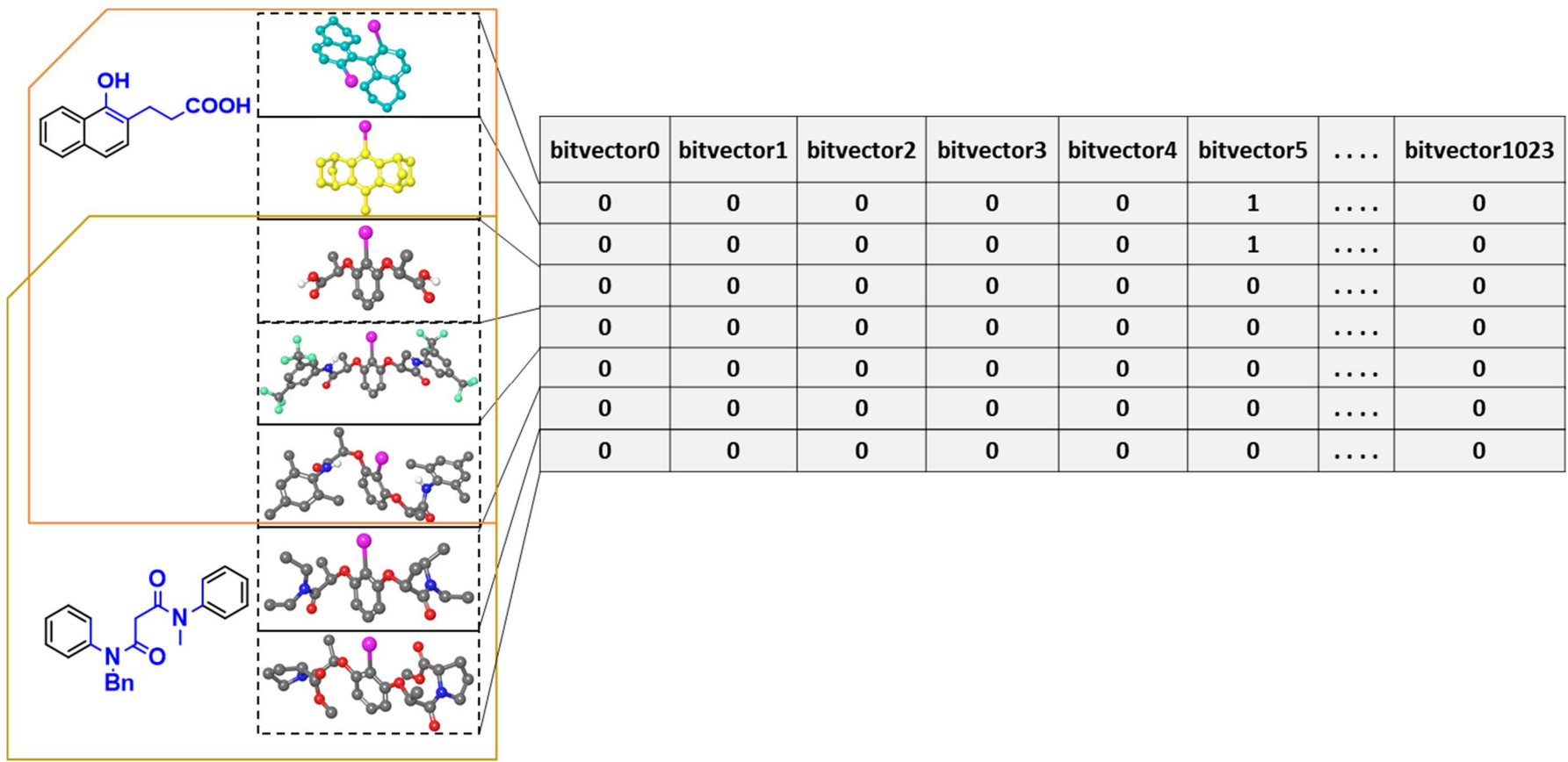

**Figure 5B.**

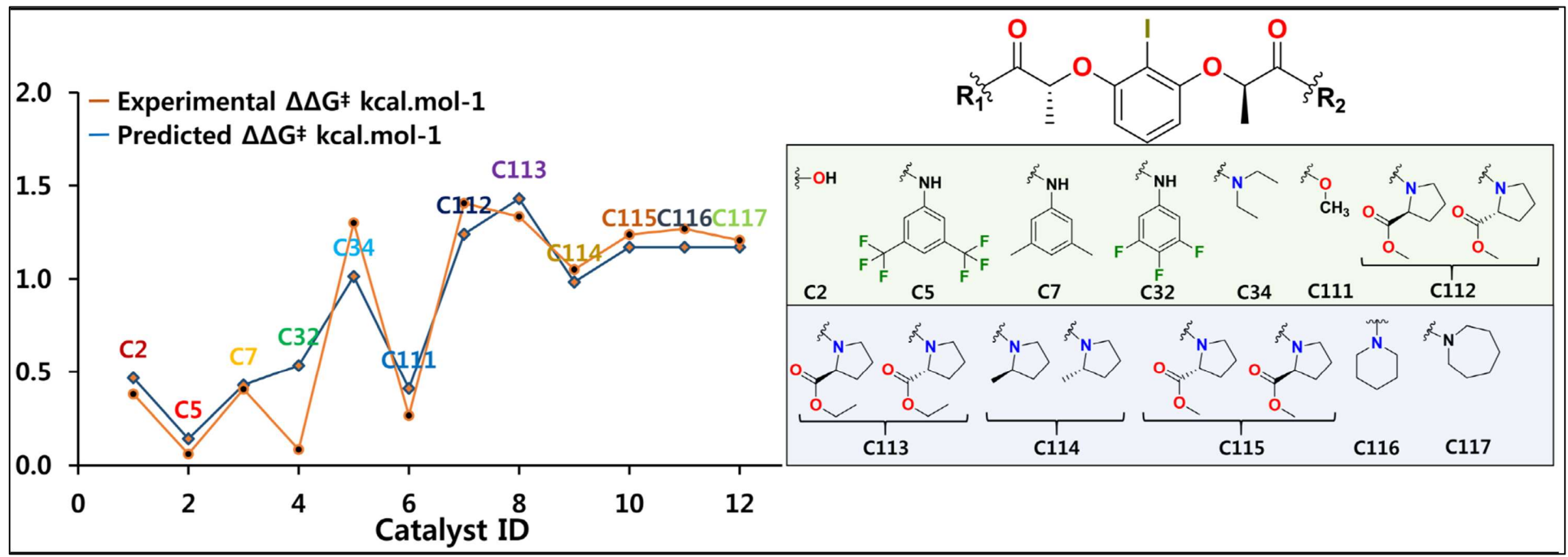

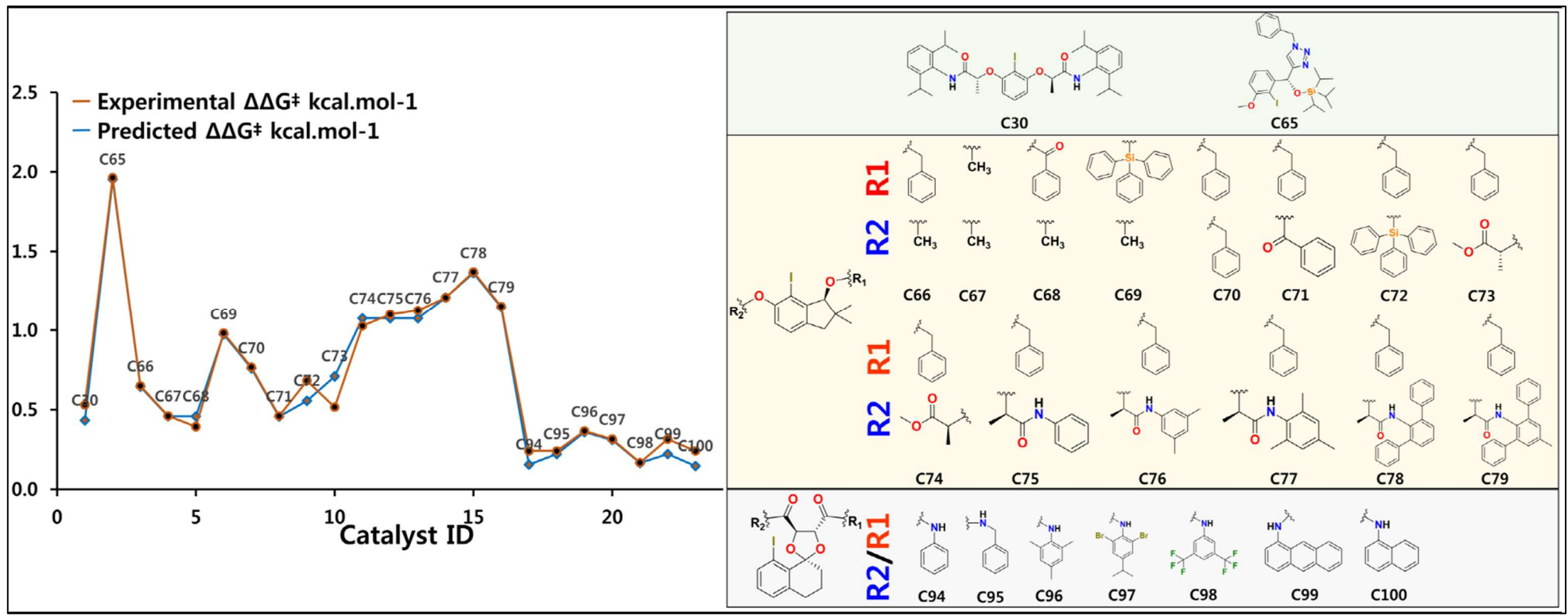

**Figure 5C.**

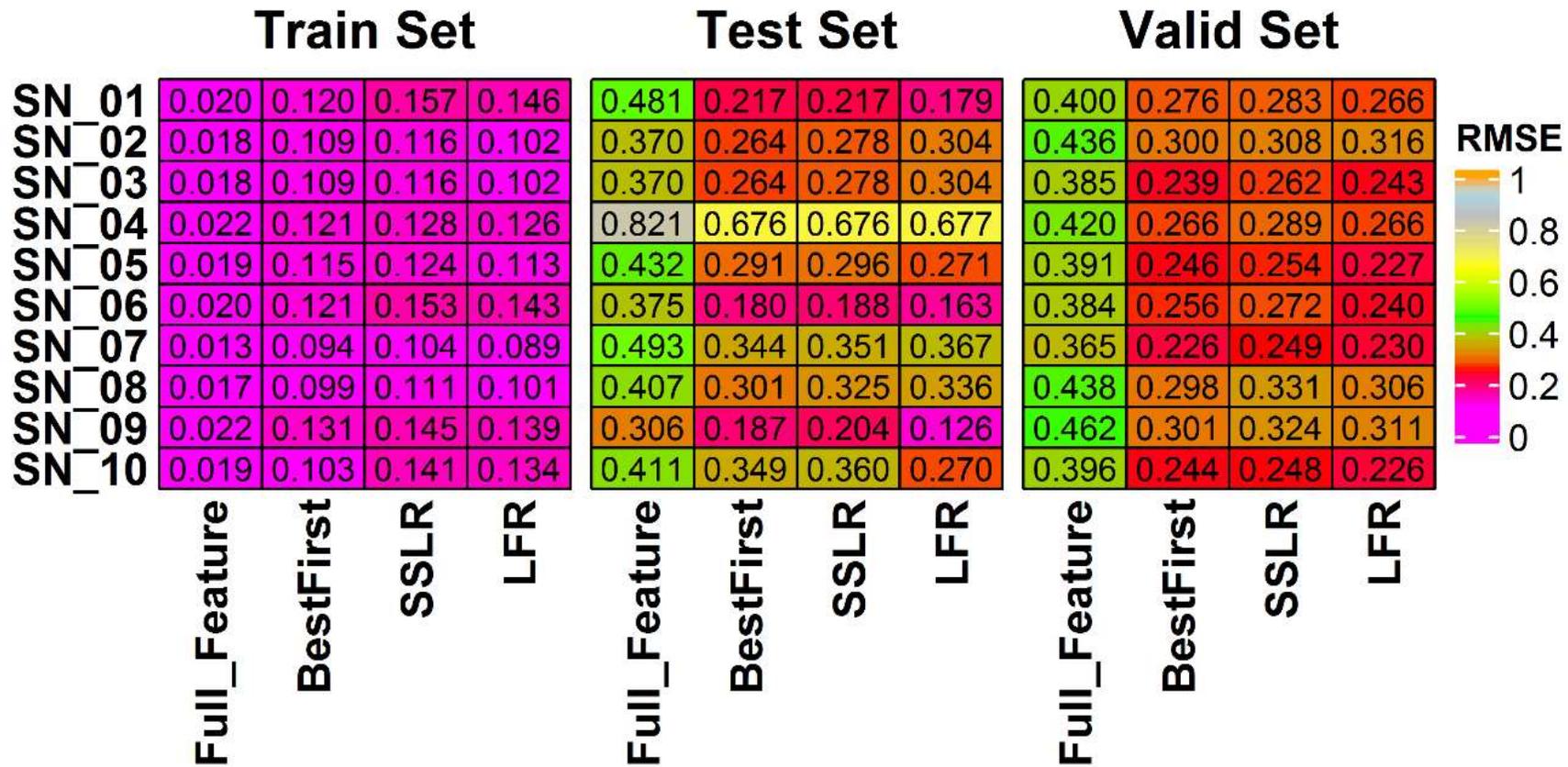

**Figure 5D.**

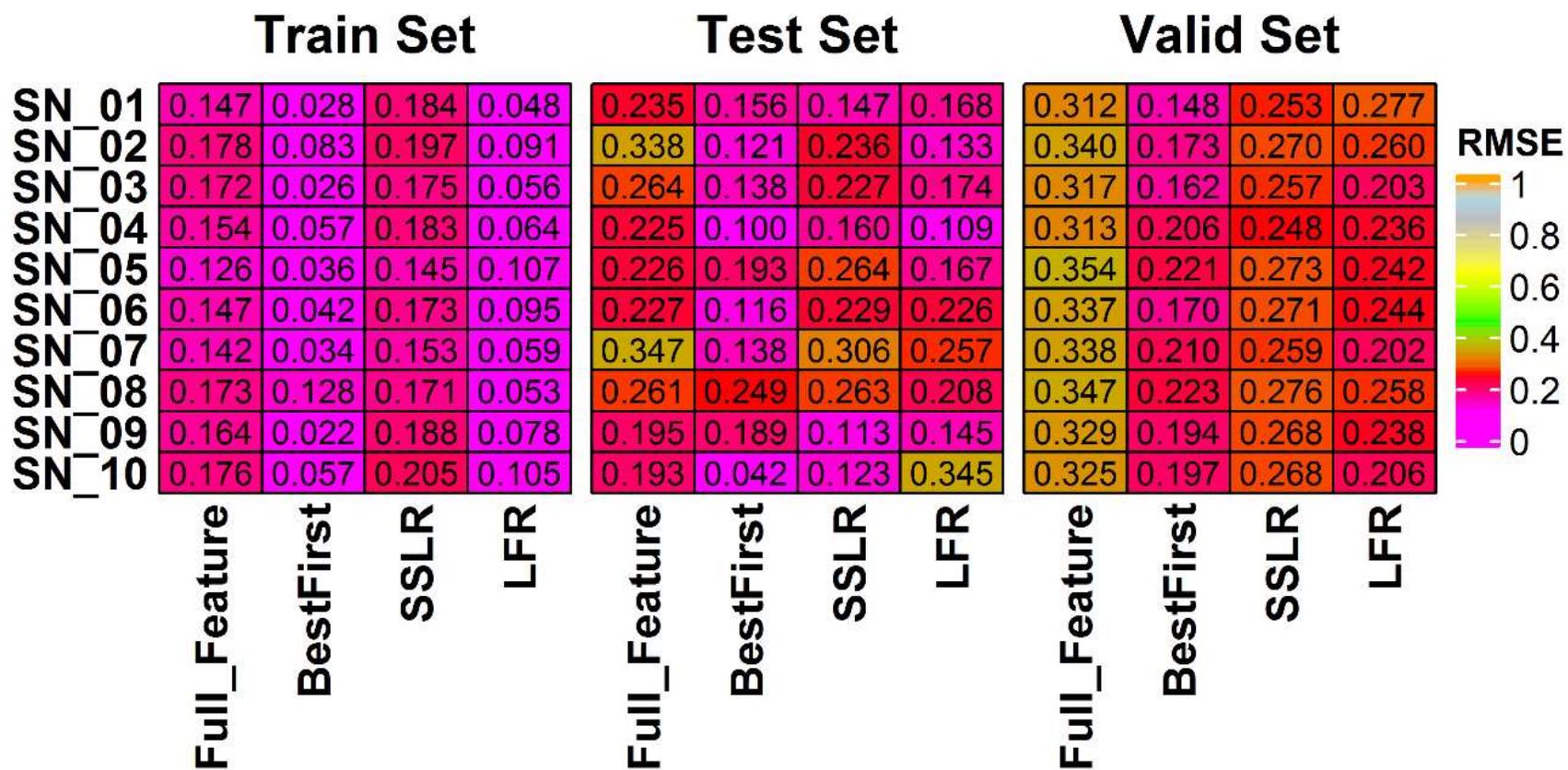

**Figure 5. Recyclability, compatibility, and generality of one hot encoded model.**

The collected data and their 3D fingerprint (E3FP) were structured, updated, and reused to build **model 2A** (enantioselective oxidative cross coupling) and **2B** (enantioselective oxidative *para* dearomatization). **(A)** structure of catalysis data & their descriptors (one hot encoded geometry) in database for general availability across name reactions. **(B)** Structure–selectivity trends according to respective catalysts (x-axis: substituents of catalysts, y-axis: selectivity in each model 2A, and 2B). **(C)** Robustness of model 2A. With four type feature selectors (Full, Best-First, SSLR, and LFR), 10 Runs of the best regressor (Gaussian Processes) consistently presented reliable determination coefficient and RMSE values for TR, TS, and 5-CV. SN means seed number of each random states. **(D)** Robustness of model 2B. With four type feature selectors (Full, Best-First, SSLR, and LFR), 10 runs of the best regressor (Random Forest). The determination coefficient and RMSE values were respectively presented for TR, TS, and 5-CV. SN means seed number of each random states.

**Figure 6A.**

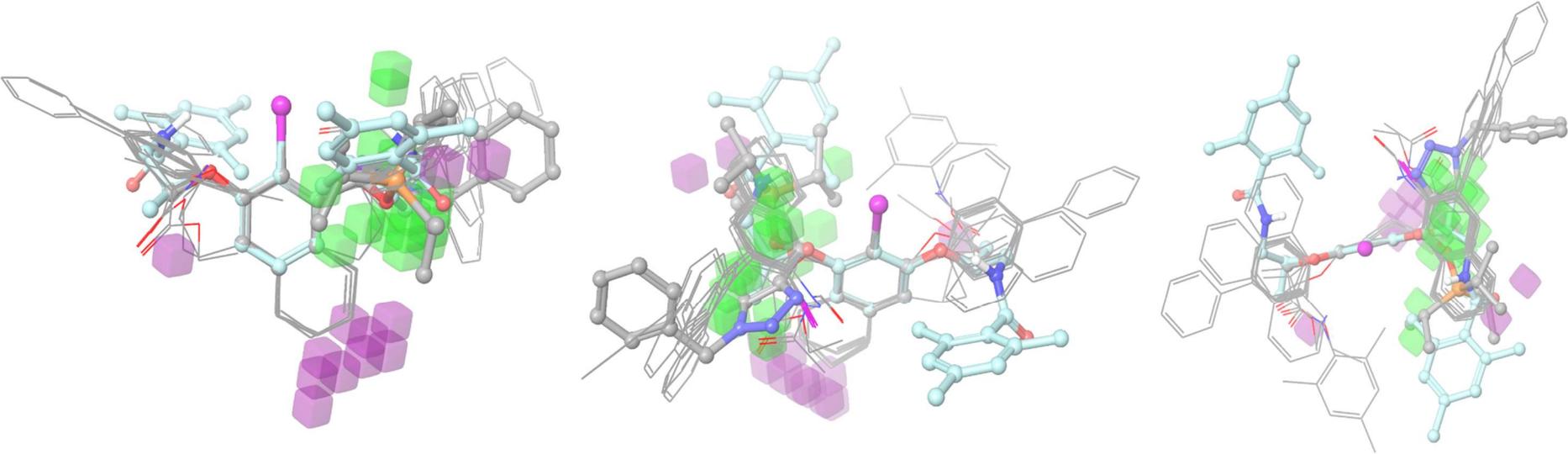

**Figure 6B.**

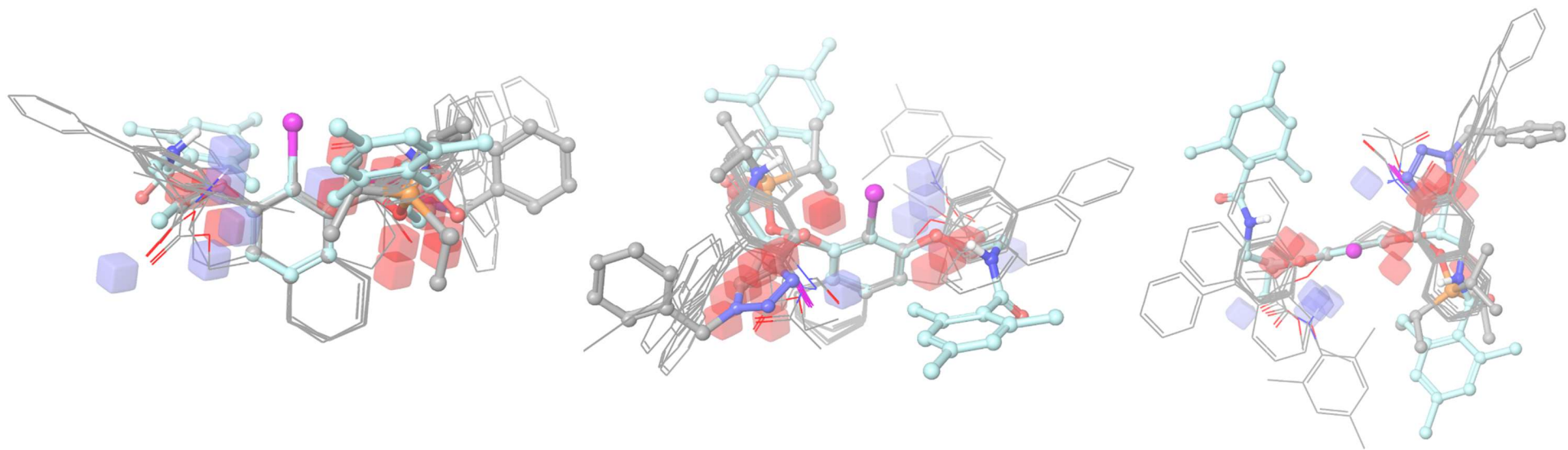

**Figure 6C.**

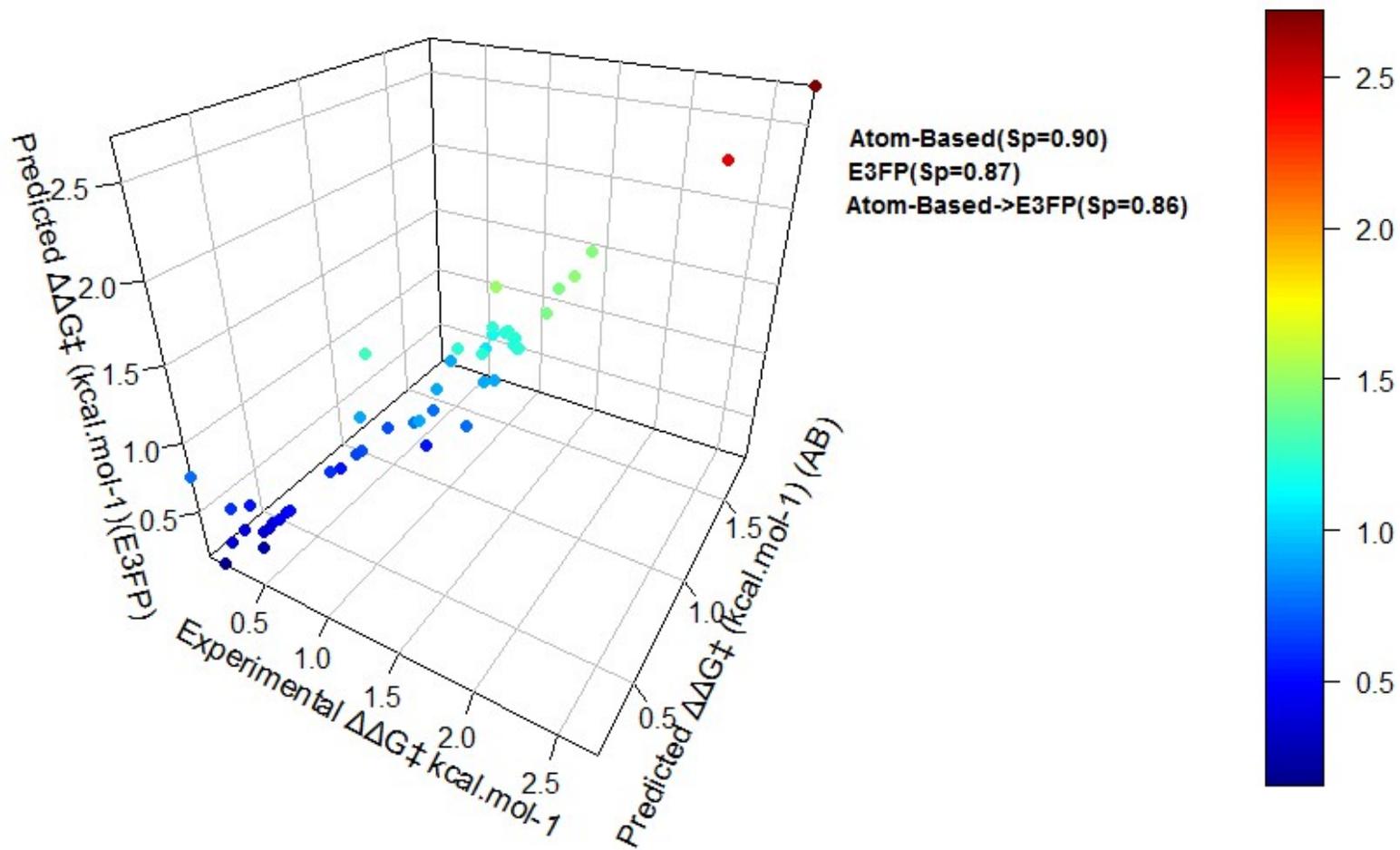

**Figure 6D.**

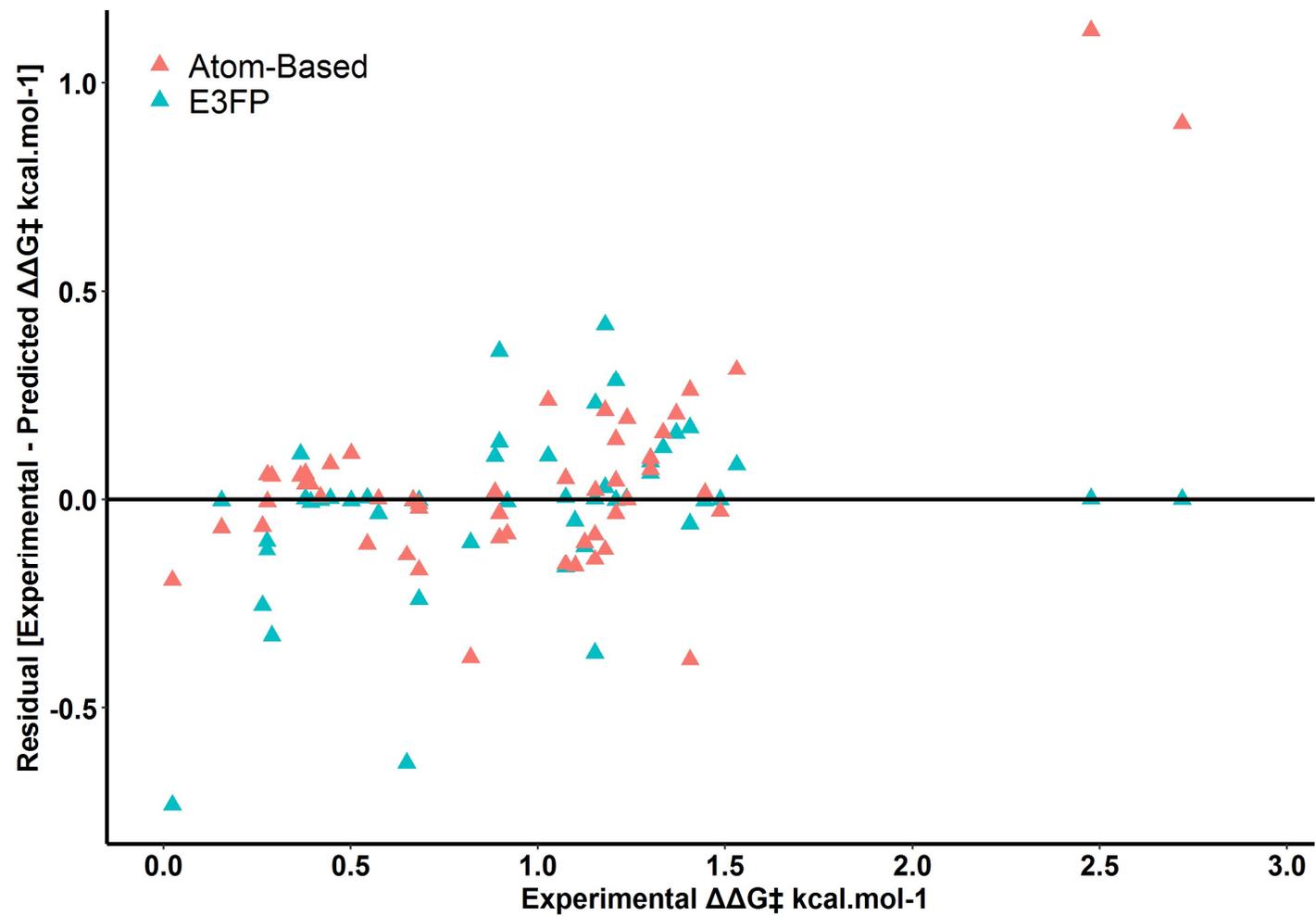

**Figure 6. The consensus prediction of the ensemble model.**

Predicted values of 3D fingerprint (E3FP) based regression models could show the consensus with NCI feature vector based prediction as well as experimental values. **(A)** The hydrophobicity contour map of atom based NCI model. Front view (left), back view (middle), and top view (right). Two highest potent catalysts were chosen for the visualization. Catalyst C13 (cyan carbon ball and stick) of Ishihara, C65 (gray carbon ball and stick) of Nachtsheim. Cubes having electron withdrawing coefficient of more than $3.15 \times 10^{-3}$(red cube) and less than $-3.15 \times 10^{-3}$(blue cube) **(B)** The electron withdrawing contour map of atom based NCI model. Cubes having hydrophobic coefficient of more than $6.5 \times 10^{-3}$(green cube) and less than $-6.5 \times 10^{-3}$(purple cube) **(C)** 3D scatter plots visualized the correlation between experimental selectivity, E3FP based predicted selectivity, and NCI based predicted selectivity. **(D)** Residual values between predicted and experimental selectivity showed that E3FP based prediction can provide more practical forecasting on high active catalysts with narrower residual values.

**Model based augmentation for asymmetric oxidative transformation via hypervalent iodines.** When researchers study the molecular recognition of any binding complex such as catalyst-substrate or protein-ligand, they hope to pinpoint a particular dominant interaction among plausible NCIs. In the case of the chiral scaffold of cluster 1 in Figure2A, helical chirality of these C2 type L-I-L arenes (eg. Ishihara, Wirth, Muniz, and Gong groups' catalysts) [18, 24, 47] can be explained with three dominant interactions: (1) fixing a dominant conformation by the intramolecular interactions (hydrogen bonding or n−σ* bonding) between amide (or ester) C=O and ligand (exceptionally iodine in Jacobsen catalyst), (2) blocking one enantiotopic face of a substrate through the optimal balance between their pi-pi interaction and steric hindrance (between substrate aryl group and catalyst aryl group), and (3) the biased arrangement of the hydrogen bonding (the removal of another chance) in C2 axis through the best torsional angle between $R^2$ (eg. chiral flag methyl group) and iodoaryl group (Figure7A). Sunoj's qualitative DFT model well described this mechanistic view with simulated geometry parameters and $\Delta\Delta G^{\ddagger}$ [27,29]. However, the insight cannot explain high enantioselectivity of simple catalysts such as tertiary alkyl amide of $R^1$ position in intramolecular cross coupling. Furthermore, because almost key interactions are common in both high and low potent catalysts, 'induced fit' model (of receptor-ligand recognition) [70] of every NCI than a dominant NCI is more suitable for delicate organocatalyst-substrate recognition. During the dynamics of this complex, it is supposed that the induced fitting with secondary interactions (smaller but the larger numbers than primary) makes the fluctuation of the initial geometry, which is arranged by primary interaction, and disrupts linear regression of dominant NCIs (chosen variables) constructed by human recognition. If the global mechanistic view (across reaction space or catalyst space) cannot be quantified through such linear model of dominant NCIs, how can it be explained quantitatively?

[Insert Figure7 here]

**Fig 7.: The interpretation of enantioselective Kita oxidative spirolactonization.**

For this purpose, this study added E3FP bitvectors (one hot encoded geometry) into known descriptors and compared their performance. While the bitvectors follows their absolute definition (radial geometry), any NCI feature follows relative definition. Thus, embedding of NCI features into numerical structures (scalar) inevitably lose a part information of DFT calculation. Furthermore, the pairwise dependency resulted in generality across name reactions. Even if E3FP based models cannot directly be translated into important geometrical parameters (angle, distance, and charge) or dominant NCI as like conventional mechanistic studies, these model provide robust and reliable response (screening & prediction) to human question (designed library). To solve this tension between the predictive power and interpretability of model predictions, feature importance analysis was conducted using dataset 1. Notably, the contribution of individual bitvectors could be understood through (1) the SHAP analysis [69], and (2) the bitvector comparison between catalysts. SHAP values of every bitvector could be achieved through the process, which consists of data oversampling from training set, prediction of sampled data, and calculation of shapley regression value under LIME model. SHAP values for respective bitvectors show which feature is more important for improving accuracy (Figure7B & 7C). In other words, this feature importance plot gives an idea, about, which features should be or not be present in a catalyst to achieve high enantioselectivity (% ee) value. A large absolute SHAP value of a chosen bitvector means that the chosen feature can improve prediction accuracy after adding the feature. The absolute size of the SHAP value is proportional to the amount of change in prediction accuracy. The +/- sign of SHAP values mean positive contribution for active catalysts and negative for inactive catalysts.

Firstly, we priori focused on bitvectors occupied in active catalysts. The Best First algorithm searched the space of feature subsets by greedy hill-climbing augmented with a backtracking facility to pick up variables for building classification model. When analyzing the best classification model (Best-First-NB) of oxidative dearomatization, bitvector180, bitvector265, bitvector331, and ~~bitvector393~~ are more prevalent in inactive catalysts (frequency>4) in the SHAP summary plot (Figure7B). Meanwhile, bitvector450, bitvector38, bitvector89, bitvector0, bitvector204, bitvector515, bitvector167, and bitvector341 were occupied by active catalysts (Figure7B). Notably, the bitvector450 and bitvector204 was the most important features for active prediction. Both bitvector204 and bitvector450 were frequently found in active catalysts, C81, C82, C83, C84, C85, C86, C87, C88, and C102 (% ee: 63 to 83). The unique SMART pattern of these catalysts was spiro[4.4]nonane ring system of spiroindene framework, which is lesser crowd around iodobenzene, to make them active (%ee: 65-84). Exceptionally, C102 doesn't have this spiro substructure but could get the bitvector 'on' through the consistent 3D geometry with both the spiro ring and its substituent. These bitvectors suggest that spiro five-membered rings need to be intensively introduced into an innovative catalyst design. Moreover, the force-plot for a potent catalyst C86 among these catalysts obviously revealed that bitvector204 and bitvector450 has maximum force (both with bitvector 'on') towards the predictability (Figure7D). In addition, the bitvector38 is present in catalysts C4, C5, C6, C13 (Train set). After SMART analysis of maximum common substructure, we found that substructure (-O-C(C)-C-) focusing around the central iodobenzene core were common among them (Supplementary Table). The force-plot for the most potent catalyst (C13: 98 %ee) gives an overview of list of bitvectors contributing to active prediction (Figure7D). The bitvector89 also showed positive signed higher SHAP values with the presence in catalyst C5, C10, C101 (Train set), and C62 (Test set) showing 73-83 % ee.

In sequence, the absolute size of SHAP values was examined to check discrepancy or consistency with bitvectors favored by active catalysts. Expectedly, bitvector331 (Top4 important of classifier) and bitvector265 were only present in inactive catalysts, C8, C47, C48, C50, C51 (for bitvector331), C47, C48, C50, C51, C53, C54, C101 (for bitvector265), and C45, C46 (for both in Test set) respectively with low enantioselectivity of 22-52 % ee except C101 (73 % ee). As shown in Figure7D, the force-plot for a low potent catalyst, C50 (22 % ee) revealed contribution of bitvector331, the most common inactive feature in the dataset (Figure7B). The unique SMART pattern of this bitvector is tetrahydronaphthalene. This force-plot of catalyst C50 also presented that, apart from bitvector331 feature, the bitvector265 and bitvector452 features force these catalysts towards low enantioselectivity (Figure7D). Furthermore, bitvector65 (C49, C52) and bitvector565 (C2, C52) showed both have high negative SHAP values on the classification model, though their frequency was very less. These inactive catalysts (2 to 43% ee) also commonly shared fused aromatic ring (naphthalene). Likewise, the bitvector180 showing fused biphenyl group was also found in such inactive catalysts, i.e. C11, C47, C48, C51, C89 (Train set), and C45, C46, C55 (Test set) with negative coefficient. More notably, the bitvector932, which is an important feature of the best regression model in Figure 7D also indicate another fused ring system of biphenyl group with C45, C46, C48, C49, C51, C52, C91 (Train set), C47, C50, C90 (Test set) of low $\Delta\Delta G^{\ddagger}$ (0.024~1.179 kcal/mol). More interestingly, this analyzed result can be also supported by the contour map of NCI vector based model. In other words, the hydrophobicity of these fused biphenyl ring systems is only favored by one side of iodobenzene and disfavored by another side in the contour map (Figure 6A). These fused biphenyl ring systems deprive any opportunity to introduce electron withdrawing group into α, β, γ – position of iodobenzene (Figure 6B).

Thirdly, common important features for both the best regressor (SSLR-SMOReg) and the best classifier (Best-First-NB) were only bitvector926, bitvector565, bitvector89, and bitvector9 outcome (Figure 7B and 7E). Expectedly, feature importance changed according to how to treat the bins of enantioselectivity (dependent variable). Outstanding feature in these common is bitvector89 with its positive outcome (Figure 7B and 7E). The bitvector89 is present in catalysts C5, C62, C101 (Train set), and C10 (Test set) with more than 73% ee values (1.099~1.406 kcal/mol). When comparing on-bit catalysts with off-bit catalysts, the common and unique SMART pattern of this bitvector is isopropyl and methyl groups without any congeneric core structure.

Finally, unique bitvector 885 of the most potent catalyst, C13 was Top1 important feature (highest positive outcome) in the best regression model (Figure7E). The bitvector885 has the lowest frequency and is found only in catalyst C13. Interestingly, even if this catalyst of the highest $\Delta\Delta G^{\ddagger}$ shares the congeneric core structure with C1 to C10 catalysts and also share the additional substructure (-CH-CH$_2$-NH-) flanked on both sides of core (Iodo-benzene) with three catalysts (C102, 105, and C107), bitvector885 exists only in catalyst C13. The force-plot for C13 revealed that bitvector885, followed by bitvector514 maximally contributing for its high enantioselectivity ($\Delta\Delta G^{\ddagger}_{(Exp)}$= 2.72 kcal/mol). Meanwhile, the force-plot for catalyst C65 (97% ee) showed the feature importance order of bitvector80 (Top4) > bitector514 (Top7), bitvectior675 (Top9), and bitvector494 (Top3) in Figure7E. Among the sixteen co-present bitvectors in both catalyst C13 and C65, bitvector494 (Top3) is only one common important feature. The most important bitvector80 for accurately predicting catalyst 65 could be only found in catalysts C7 and C65 with ΔΔG‡(Exp) of 1.487 kcal/mol and 2.477 kcal/mol respectively. Notably, Figure 7F summarized common SMART pattern and their predicted activity of these described features and their representative catalysts.

**Figure 7A.**

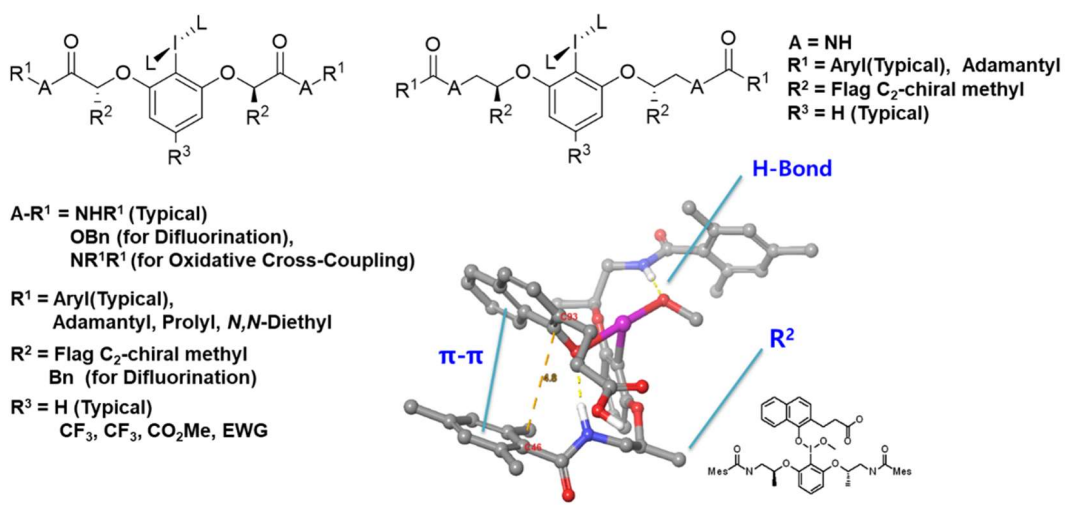

**Figure 7B.**

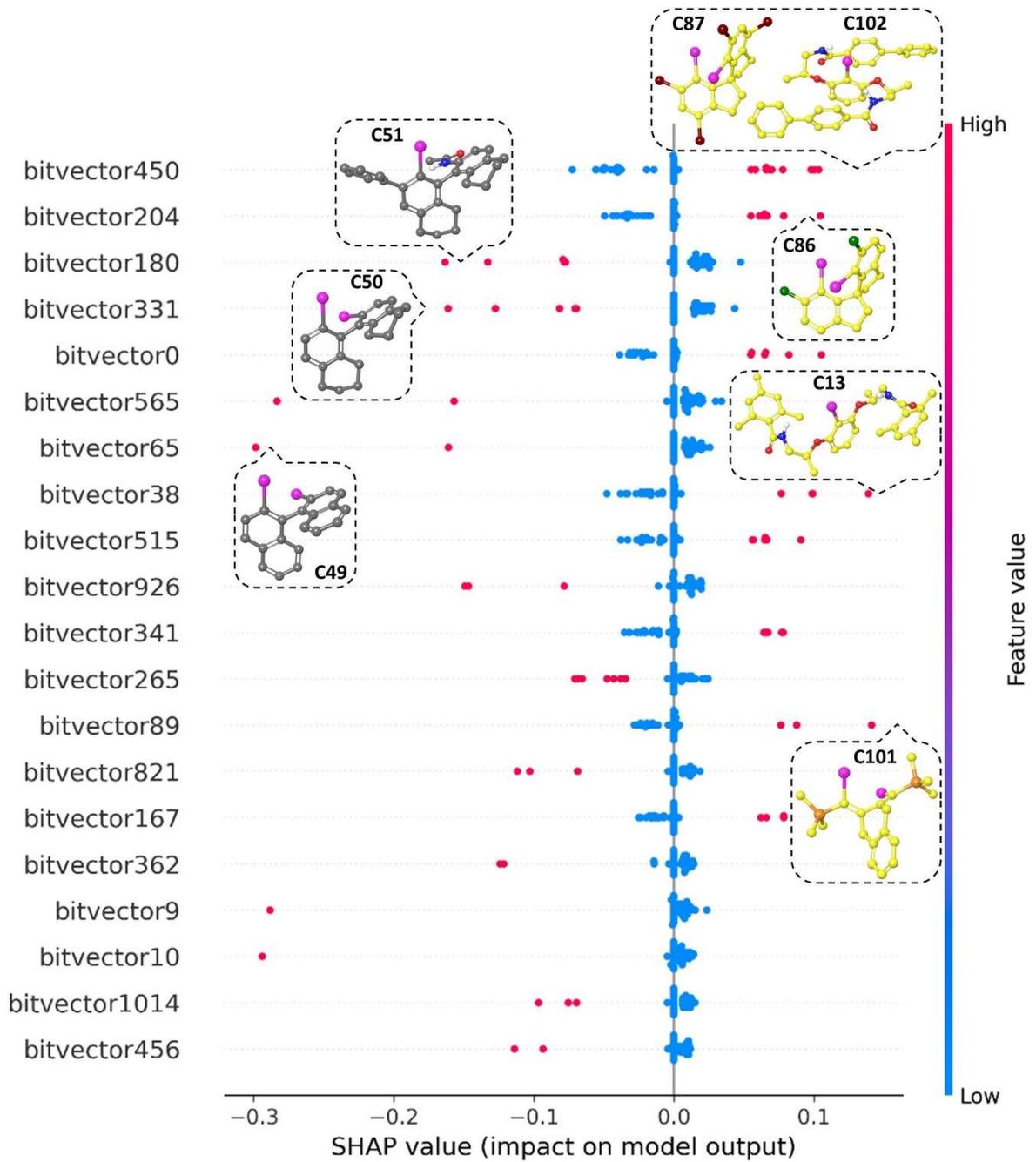

**Figure 7C.**

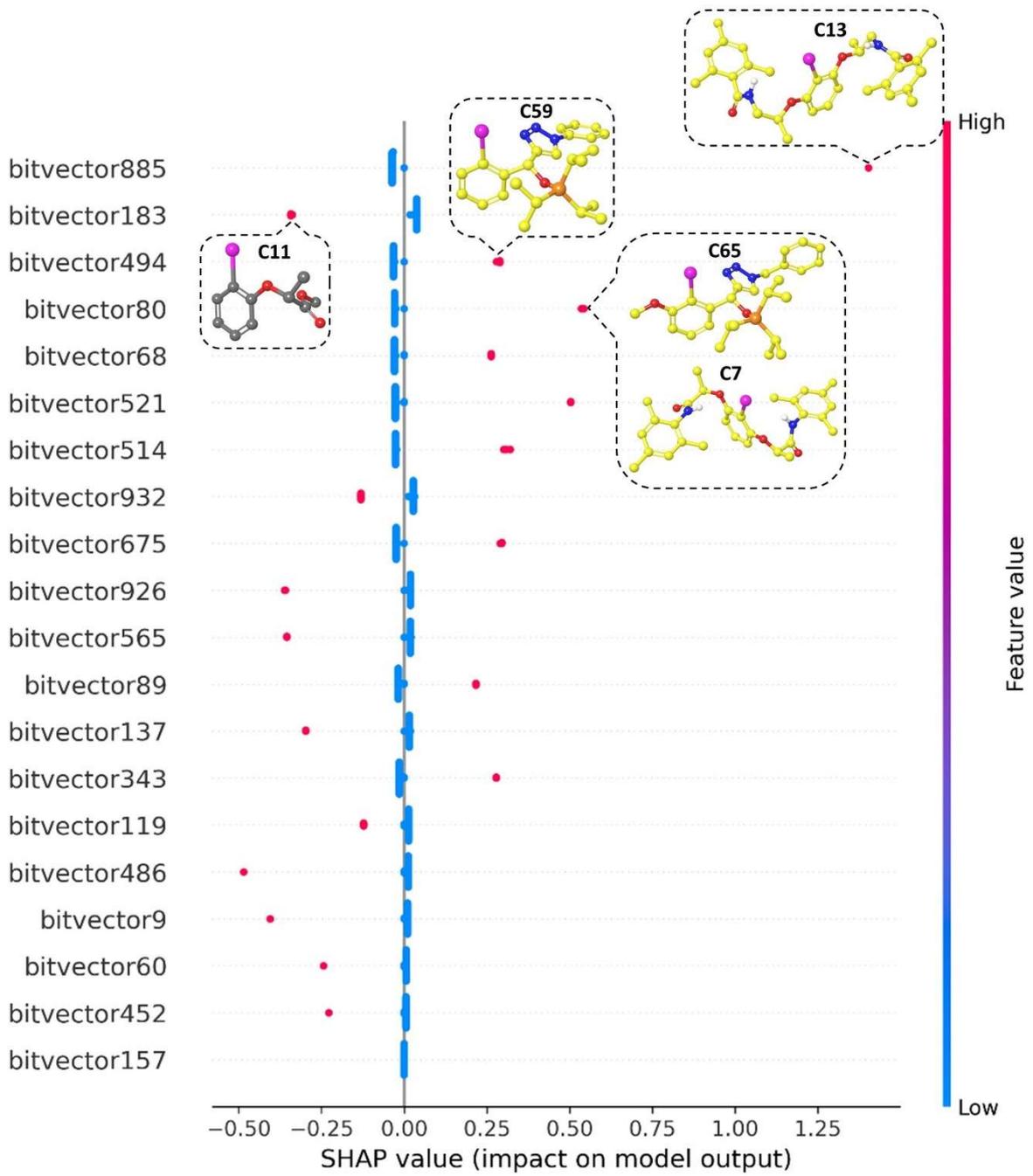

**Figure 7D.**

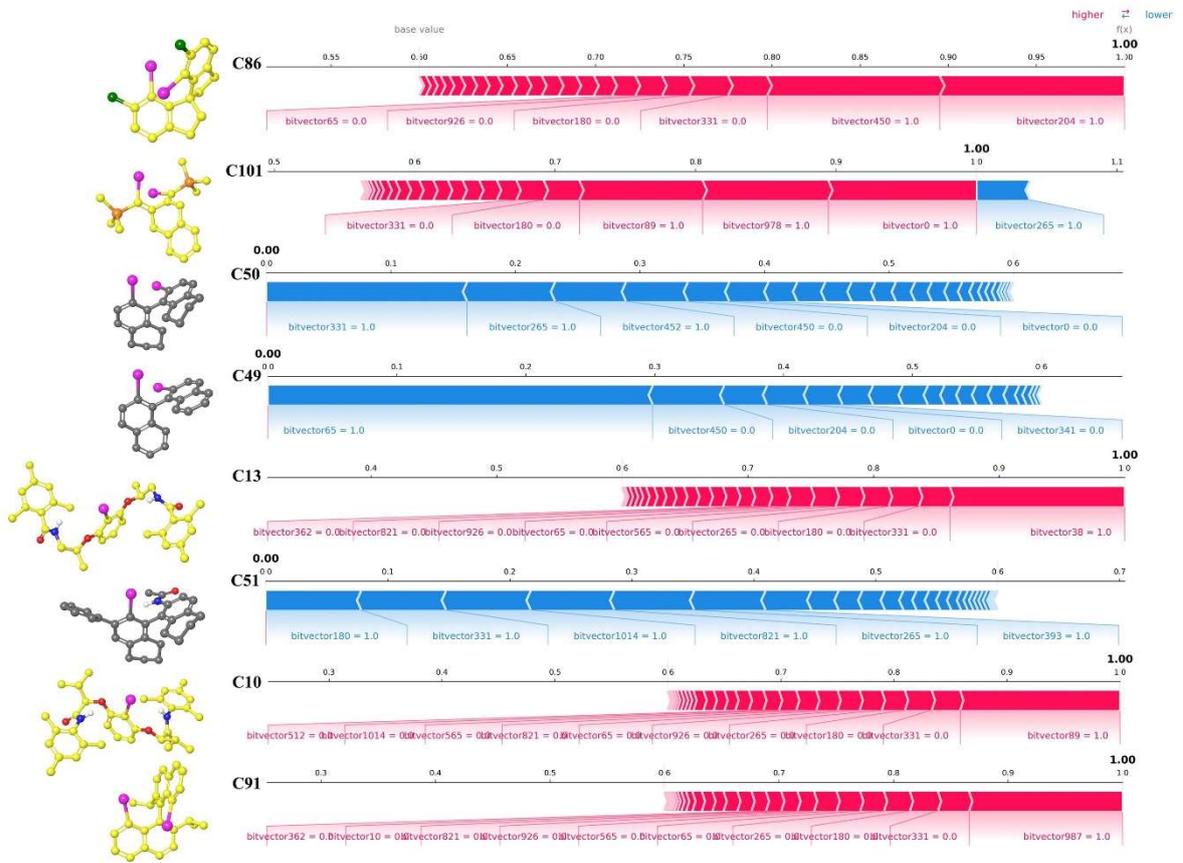

**Figure 7E.**

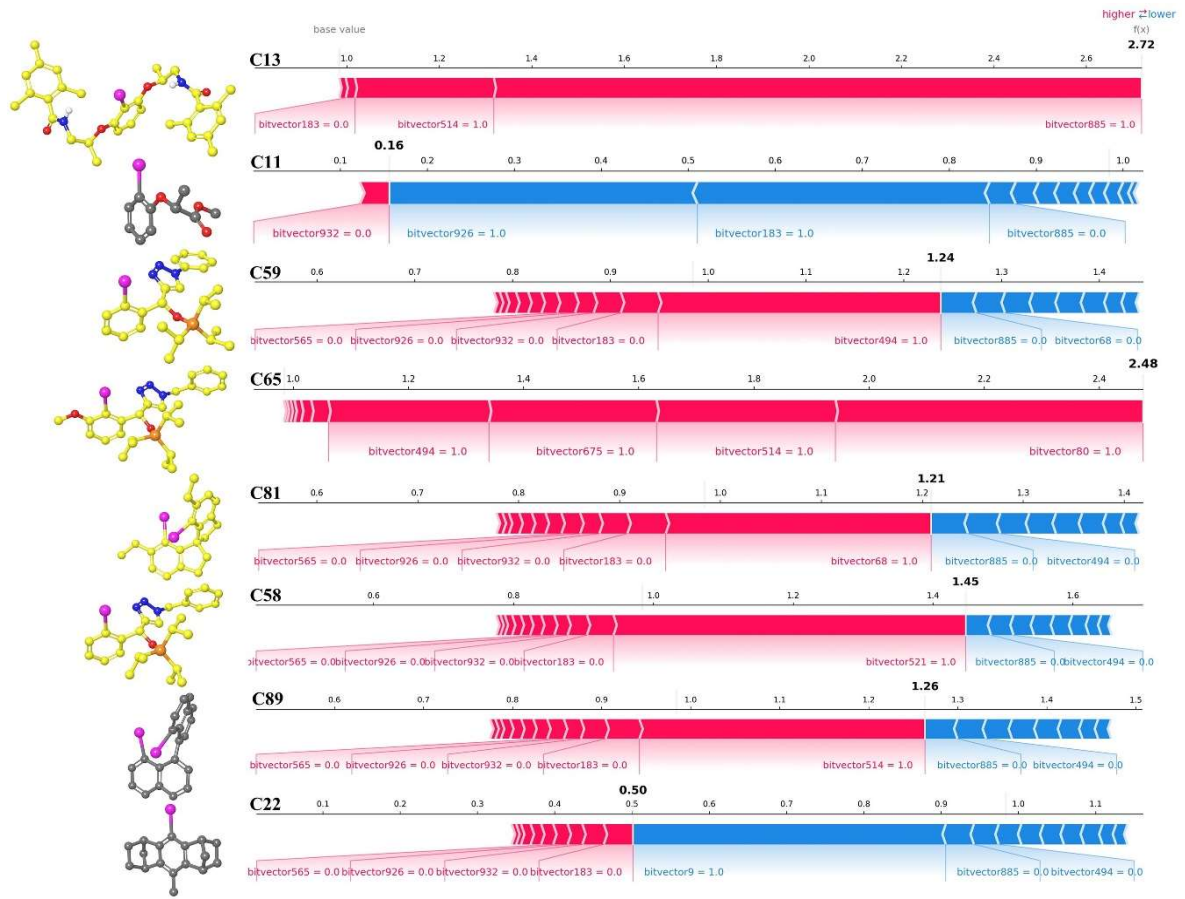

**Figure 7F.**

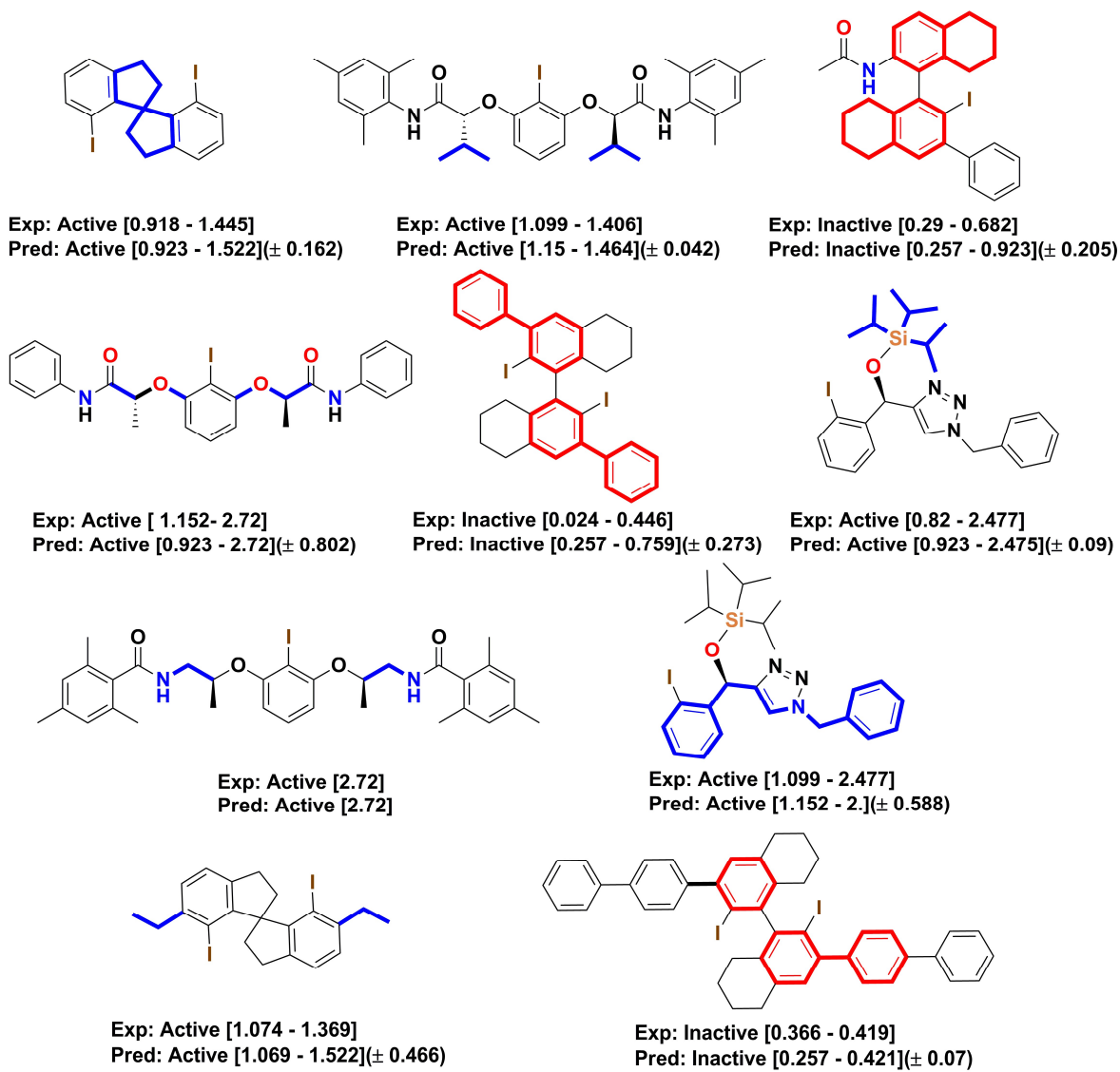

**Figure 7. The interpretation of enantioselective Kita oxidative spirolactonization.** Interpretation of 3D fingerprint (E3FP) based model. **(A)** dominant NCI based mechanistic summary before this study. **(B)** Feature importance analysis of the E3FP based best classifier (Best-First-NB). **(C)** Feature importance analysis of the E3FP based best regressor (SSLR-SMOReg). **(D)** SHAP values (force plot with predicted activity) of each chosen catalysts in Best-First-NB model. **(E)** SHAP values (force plot with predicted activity) of each chosen catalysts in SSLR-SMOReg model. **(F)** Schematic representation the selected precatalyst. The bitvectors representing the substructures are highlighted (Red color and blue color indicates

the negative and positive impact on the overall predictability of ML model). The range of experimental and predicted enantioselectivity values were mentioned for precatalysts containing respective substructures. The values in parenthesis represent the std. dev. from the residual from experimental and predicted values.

**Conclusion**

In summary, enantioselective prediction of chiral iodoarenes (III) could be quantified with high statistical performance ($Q^2$, $R^2$, and RMSE). The classification and regression models built based on the stereoselectivity data of three name reactions encourages that further investigation with respect to application beyond the studied reactions. Moreover, we recommended ensemble prediction guided by three type molecular descriptors, (1) one hot encoded radial geometry (E3FP bit vectors), (2) local steric parameters (sterimol), and (3) NCI vectors. Before our study, an experiment was a more efficient way of knowing the reaction scope of chiral iodoarenes than computation in terms of time cost. Now we believe that our workflow and models can make synthetic chemists efficiently predict potency of their catalysts and select them for their synthesis from their designed catalysts. We believe that our workflow and models can help synthetic chemists to efficiently predict the potency of their designed catalysts and selectively synthesize more promising catalysts. More notably, the generality, recyclability, and compatibility of our featurization will result in a lower computational cost than descriptive models on transition states and their energy gap. Furthermore, our method can be applied to name reactions involving unsolved transition states.

**Methods**

**Data collection & DFT calculation.** Data were collected from literature and X-Ray

diffraction data of Cambridge Crystallographic Data Center (CCDC ID: 1404588, 830345, 1570404, 1495386, and 1040056). Optimization calculations are used to acquire the coordinates of the atoms of the molecule, where it has the minimum energy and is hence stable. Quantum chemical calculations were carried out with the Gaussian 09 suite [71]. In this work, the geometry optimization of each pre-catalysts and catalysts structure was optimized to respective minimum energy geometries using the hybrid-meta GGA functional M06-2X method with a mixed basis set LanL2DZ for I and 6-31G++(d,p) for other atoms [72]. The optimized geometry was validated through RMSD comparison with the CCDC data and the absence of imaginary frequencies. M06-2X at 6-31G++(d,p) was selected based on its successful performance and wide usage in the literature on hypervalent iodine compounds for different systems [27, 29, 32]. The SMD implicit solvation model was used to account for solvation effects based on the respective reaction conditions [73-74]. To determine whether the geometry optimization has found a minimum, frequency calculations were performed at the same computational level. Single-point energy calculations were performed at the PCM-M06-2X/6–31G** with LanL2DZ for the I level of theory. Therefore, the starting data was an X-ray structure with their availability whenever possible and obtained geometry optimized structures.

**Descriptor Generation.** 3D fingerprints (E3FP) [39] and sterimol parameters [75], describing the global and local environment of molecular structures, were generated from respective optimized geometry. In brief, for given molecular geometry optimized through DFT calculation, the RDKit library, an open-source python library, produced the 3D fingerprints in a bit-vector form, where a specific bit indicates the presence of a 3D molecular substructure regardless of covalent bonding or not. 3D fingerprints of 1024 bit-vector were split and generated 1,024 independent variables. To validate captured their 3D information, multiple conformers of iodoarenes were encoded into 3D fingerprints and 3D similarity of conformer pairs was calculated for checking their duplicated encoding or discriminative power of this molecular representation. Furthermore, the 3D similarity distribution was also compared with maximally overplayed ROCS similarity, which was calculated from Gaussian functional molecular shape of the conformers by OE Toolkit (center of mass, translation and rotation) in supplementary information S. Figure1. For sterimol descriptors, to distinguish different type of steric interactions features represented by L, B1 and B5, which describes the maximum length, minimum width and maximum width respectively for a given position in the molecule were calculated according to method of Sigman group. NCI feature vectors were generated from respective optimized geometry by Phase and Field based QSAR (Schrodinger Suit 2018-4). The generated features were selected based on statistical performance (R squared of TR, CV, and TS) of their PLS models. In the contrast to 3D fingerprints (or sterimol parameters), these NCI features could not be recycled due to their pairwise property and alignment dependency.

**Machine learning methods and attribute selection.** In this work, for classification and regression a set of machine learning methods, such as Random Forest (RF) [60], Decision Tree (DT) [61], Naïve Bayes (NB), Gradient Boost Tree (GB) [63], Xtra-Gradient Boost Linear (XGB-Linear), Xtra-Gradient Boost Tree Linear (XGB-Tree) [64], Weka-Linear, Support Vector Machine (SMOReg) [67], Gaussian Processes (GP) [68], and Multi-Layer Perceptron (MLP) [76] were used to correctly identify and establish a correlation between global and local feature with experimentally determined enantioselectivity of catalyst. Here, for classification and regression ML method, we build KNIME Workflow using the KNIME Analytics platform [77]. For features/descriptors selection, we used an automatic attribute selection methods (CfsSubsetEval), combined with Best-First (BF) [78], Subset Size Forward Linear Regression

(SSLR), and Linear Forward Regression (LFR) as implemented in WEKA data mining software.

**Model evaluation and validation.** For any Classification or regression methods, the quality of any ML models can be evaluated on different metrics. In this work, for classification ML models, the standard evaluation metrics were derived from confusion matrix (true positive (TP), true negative (TN), false positive (FP) and false-negative (FN)). The predictability of the model was evaluated from sensitivity, specificity, MCC (Matthew's Correlation Coefficient), Accuracy, and AUC (Area under the Curve). For regression ML models, the various evaluation metrics were considered such as person correlation coefficient (PCC), coefficient of determination ($R^2$), spearman ranking correlation (Sp)/ranking power, mean absolute deviation (MAD), mean absolute error (MAE), root-mean-square error (RMSE), and p-value (significance level).

$$R^2 = 1 - \frac{\sum (y_i - \hat{y})^2}{\sum (y_i - \bar{y})^2}$$

$$Sp = 1 - \frac{6 \sum D^2}{n(n^2 - 1)}$$

$$MAE = \frac{1}{N} \sum_{i=1}^{N} |y_i - \hat{y}|$$

$$RMSE = \sqrt{\frac{1}{N} \sum_{i=1}^{N} (y_i - \hat{y})^2}$$

$$MAD = \frac{1}{N} \sum_{i=1}^{N} |x_i - m|$$

Where, $\hat{y}$ is the predicted value, and $\bar{y}$ is the mean value of ΔΔG‡. $x_i$ is the $i^{th}$ number of sample and m is the mean of the sample.

We considered the MCC as a measure to evaluate the quality of the model in the classification ML method, as it is considered as a balanced measure even if the classes are of different sizes. The MCC values above 0.4 can be considered as predictive in ML methods [79].

All the classification and regression ML models were evaluated by the cross-validation method using K-fold cross-validation [80]. In general, for K-fold cross-validation, the whole samples are divided into K subsamples. From the K subsamples, a single subsample is retained as a valid set to test the model, and the rest K-1 subsamples are used as a train set. This procedure is repeated K times with each K subsamples used once as a valid set and cross-validation metrics are calculated. Here, all the cross-validation was performed with 5-fold cross-validation under the KNIME Analytic Platform.

**Data availability**

All data relating to this study is available in the Supplementary Information.

**Conflict of interest**

The author(s) confirm that this article content has no conflicts of interest.

**Acknowledgment**

The acknowledgment will be added after the blind peer review.

**Author contributions**

The contribution will be added after the blind peer review.